# Structural and dynamical investigation of glassforming smectogen by X-ray diffraction and infra-red spectroscopy aided by density functional theory calculations

Aleksandra Deptuch[1,*], Natalia Górska[2], Stanisław Baran[3], Magdalena Urbańska[4]

[1] Institute of Nuclear Physics Polish Academy of Sciences, Radzikowskiego 152, PL-31342 Kraków, Poland

[2] Faculty of Chemistry, Jagiellonian University, Gronostajowa 2, PL-30387 Kraków, Poland

[3] Jagiellonian University, Faculty of Physics, Astronomy and Applied Computer Science, M. Smoluchowski Institute of Physics, Łojasiewicza 11, PL-30-348 Kraków, Poland

[4] Institute of Chemistry, Military University of Technology, Kaliskiego 2, PL-00908 Warsaw, Poland

[*]corresponding author, aleksandra.deptuch@ifj.edu.pl

## Abstract

Molecular arrangement in the chiral smectic phases of the glassforming (S)-4'-(1-methylheptylcarbonyl)biphenyl-4-yl 4-[7-(2,2,3,3,4,4,4-heptafluorobutoxy) heptyl-1-oxy]benzoate is investigated by X-ray diffraction. An increased correlation length of the positional short-range order in the supercooled state agrees with the previous assumption of the hexatic smectic phase. However, the registered X-ray diffraction patterns are not typical for the hexatic phases. Comparison of the smectic layer spacing and optical tilt angle indicates a strongly non-linear shape of molecules, which enables choice of the molecular models obtained by DFT calculations, used subsequently to interpret the infra-red spectra. The presumption of the hexatic smectic $F_A$* or $I_A$* phase is supported by the splitting of the absorption bands related to the C=O stretching in the supercooled state, which is absent in the smectic $C_A$* phase above the melting temperature. The glass transition affects the temperature dependence of the smectic layer spacing but only subtly impacts the infra-red spectra. Application of the k-means cluster analysis enables distinction between the infra-red spectra below and above the glass transition temperature, but only for certain spectral ranges.

## 1. Introduction

The family of chiral, partially fluorinated liquid crystalline compounds, with the general formula presented in Figure 1, was introduced in 2008 [1] and extended in 2011 [2] and 2015 [3]. The abbreviation used in this paper is $mX_1X_2n$, where m is the length of the $C_mH_{2m}$ chain, $X_1$, $X_2$ = H or F describe the number and position of the fluorine atoms substituted to the benzene ring and n is the length of the $C_nH_{2n+1}$ chain. The liquid crystalline phases exhibited by the $mX_1X_2n$ compounds are the smectic phases characterized by the layer order (Figure 2): paraelectric smectic A* (SmA*), ferroelectric SmC*, antiferroelectric $SmC_A$* and hexatic $SmX_A$* ($SmF_A$* or $SmI_A$* – they cannot be distinguished based on the available data) [1-6].

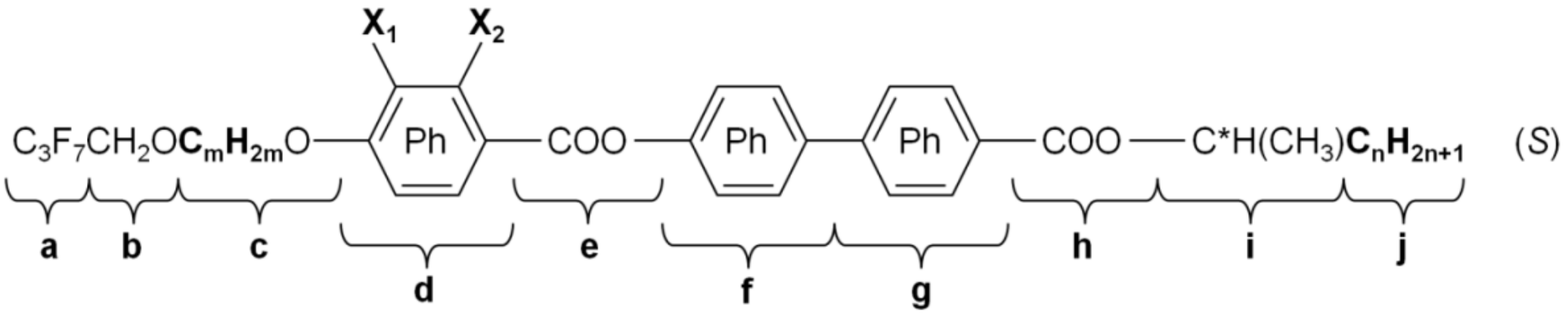

Figure 1. The general molecular formula of the $mX_1X_2n$ compounds. The notations of the particular fragments of a molecule are used in the description of the intra-molecular vibrations.

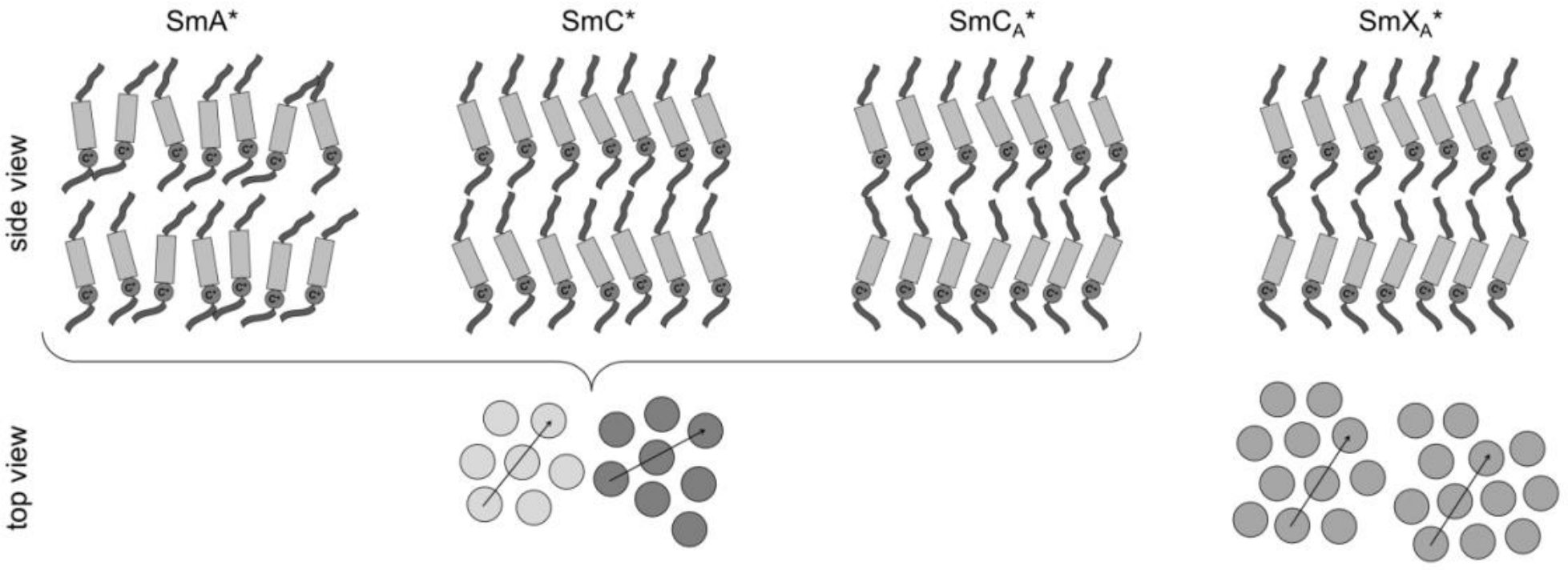

Figure 2. Schemes of the molecular arrangement in the selected smectic phases.

In the SmA* phase, the average tilt $\Theta_O$ of molecules within the smectic layers is equal to zero, while in the SmC*, SmC$_A$* and SmX$_A$* phases, $\Theta_O > 0$ [7]. The subscript "O" refers to "optical" and indicates the tilt angle measured by the electro-optic (Clark-Lagerwall) method [8]. The optical tilt angle $\Theta_O$, describing the tilt of the molecular cores, may be different than the steric tilt angle $\Theta_S$, calculated from the molecular length and the smectic layer spacing [9-13]. The order of the tilt direction in the SmC*, and SmC$_A$* phases is approximately synclinic and anticlinic, respectively [14,15]. Small deviations from the ideal syn- or anticlinic order are caused by an additional helical change in the tilt direction, with the helical pitch at the order of 100-1000 nm [2,4]. The intra-layer positional order of molecules in the SmA*, SmC* and SmC$_A$* phases is the short-range one. The hexatic SmX$_A$* phase also shows the bond-orientational order, which means that all domains with a local hexagonal ordering are oriented in the same direction within the smectic layer [7].

The tilt angle $\Theta_O$, measured in the SmC$_A$* phase for the $mX_1X_2n$ compounds and their mixtures, is ~45° [2,4,16]. It is a desired property for an antiferroelectric liquid crystal intended for application in displays because it prevents light leakage in the dark state of a display [17-19]. Another interesting property is that in some $mX_1X_2n$ compounds, the glass transition (vitrification) of the SmC$_A$* or hexatic SmX$_A$* phase is detected, which is followed by the cold crystallization after reheating above the glass transition temperature [5,6,20-23]. Glassformers undergoing the cold crystallization may serve as materials for energy storage [24,25].

The subject of this paper is the 7HH6 compound, which forms the glass of the hexatic $SmX_A$* phase. The phase sequence of 7HH6 during heating after vitrification of the smectic phase is gl$SmX_A$* (229 K) $SmX_A$* (252 K) Cr2 (271 K) Cr1 (329 K) $SmC_A$* (390 K) SmC* (394 K) SmA* (396 K) Iso, [23]. The purpose of the present study is the application of the X-ray diffraction (XRD) method to confirm the reported phase sequence and to determine the structural parameters of the smectic phases. The temperature dependence of the smectic layer spacing and the molecular models obtained by density-functional theory (DFT) calculations are applied to obtain the tilt angle of molecules and compare it with the values determined earlier by the electro-optic method [16]. The most probable molecular conformations are selected based on comparing the tilt angle determined by two different methods. Subsequently, these molecular models are used to calculate the vibrational modes and to interpret the experimental infra-red (IR) spectra. The IR spectroscopy results obtained for various temperatures serve to investigate the changes in the intra-molecular vibrations upon the glass transition, cold crystallization, melting and transition to the isotropic liquid phase of the 7HH6 compound.

## 2. Experimental and computational details

The synthetic route of (S)-4'-(1-methylheptylcarbonyl)biphenyl-4-yl 4-[5-(2,2,3,3,4,4,4-heptafluorobutoxy) heptyl-1-oxy]benzoate, abbreviated herein as 7HH6, is described in [1,2].

The X-ray diffraction measurements were carried out with the Empyrean 2 (PANalytical) diffractometer with the Cryostream 700 (Oxford Cryosystems) temperature attachment. The sample was heated to the isotropic liquid phase and placed in the glass capillary with a 0.3 mm diameter using the capillary effect. After inserting into the diffractometer, the sample was heated to 363 K, to the $SmC_A$* phase, and cooled down to 183 K with the 6 K/min rate. The patterns were collected in the geometry of the horizontal rotating capillary, in the angular range 2θ = 2-30°, upon heating from 183 K to 403 K. The programs used for data analysis were WinPLOTR [26] and OriginPro.

The FT-IR measurements were performed with the Bruker VERTEX 70v FT-IR spectrometer with Advanced Research System DE-202A cryostat and ARS-2HW compressor. The sample was mixed with KBr. The spectra were collected in the 450-4000 $cm^{-1}$ range with the 2 $cm^{-1}$ resolution and 32 scans per spectrum. The measurements were done in a vacuum upon the 1st heating of the pristine sample from 293 K to 408 K. Then the sample was cooled down directly to 173 K at 3 K/min and measurements were done on the 2nd heating to 408 K. The data analysis, including the peak fitting and k-means cluster analysis [27], was done in OriginPro.

The optimization of the molecular models and calculation of the vibrational modes were done using the Gaussian 16 program [28]. The selected basis set was def2TZVPP [29], and the exchange-correlation functional was B3LYP-D3(BJ) [30-33]. The Avogadro program [34] was used to prepare models and their visualization.

## 3. Results and discussion

### *3.1. Structural changes investigated by X-ray diffraction*

The XRD pattern of the 7HH6 sample cooled down to 183 K with the 6 K/min rate consists of a sharp peak at 2θ ≈ 5.1° and a wide diffuse maximum at 2θ ≈ 20° (Figure 3a). Subtraction of the isotropic liquid pattern at low angles also reveals another peak at 2θ ≈ 2.6°. Two low-angle diffraction peaks arise from the smectic layer order with the layer spacing $d$ obtainable from the Bragg equation $l\lambda = 2d\sin\theta$ [35], where $l$ is the peak order (1st and 2nd for 2θ ≈ 2.6° and 5.1°) and $\lambda$ is the CuKα wavelength. The cold crystallization is observed at 253 K, although the fraction of the crystal phase is very small. At 263 K, a new low-angle diffraction peak at 2θ ≈ 2.3° arises, corresponding to the layer spacing in the crystal phase. The diffraction peaks from the crystal phase become sharper upon further heating, but no transition in the solid state is observed. Two crystal phases were observed in the differential scanning calorimetry experiments performed during constant heating at 2-20 K/min rates [23]. However, in the XRD experiment, the sample crystallizes directly into the high-temperature crystal phase. The 1st, 2nd, and 3rd order peaks from the layer spacing in the crystal phase were used to calculate the $d$ value by fitting the formula:

$$\theta_l = \theta_0 + \arcsin\left(\frac{l\lambda}{d}\right) \quad (1)$$

to the plot of the peak positions vs. the peak order ($l$ = 1, 2, 3). The fitting includes the shift in the peak positions $\theta_0$, for correction of the systematic shift in $d$. The $\theta_0 \approx -0.004°$ value determined for the crystal phase was also applied to the smectic phases, where only one or two diffraction peaks related to the smectic layer spacing were visible. The XRD patterns of the $SmC_A*$ phase are similar to those in the supercooled smectic phase. However, the diffuse maximum at higher angles is wider. Only the diffuse maximum is visible in the isotropic liquid, while no low-angle peaks are observed.

In the 183-263 K range, the smectic layer spacing has a maximum of 35.0(1) Å at 233 K (Figure 3b), which is close to the glass transition temperature $T_g$ = 229 K determined from the dielectric spectra of 7HH6 [23]. In the vitrified state, $d$ increases on heating with a slope of 0.011(1) Å/K, while above $T_g$ there is a decrease of $d$ with a slope of −0.017(2) Å/K. The change in the temperature dependence of the characteristic distance at $T_g$ was also reported, e.g., for cyclo-octanol [36]. Upon the cold crystallization, the layer spacing increases by almost 3 Å. In the crystal phase, $d$ is independent of temperature, and its mean value is 37.56(2) Å. After melting, the layer spacing drops to 32.5 Å, smaller than 34.5 Å during the cold crystallization. The $d$ value increases with increasing temperature until the transition to the SmA* phase, where $d$ is approximately constant, with the mean value 36.4(6) Å. The $SmC_A*$ → SmC* transition is weakly visible as an increase in $d$ between 388 K and 389 K, while at the SmC* → SmA* transition between 394 K and 395 K, the smectic layer spacing increases by 5.4% (if $d_{SmA^*}$ is considered as 100%). It is too much to classify the SmA* phase of 7HH6 as the de Vries phase with a large tilt of molecules in random directions [37,38].

The ratio of the integrated intensities $I_{002}$ and $I_{001}$ of the 2nd and 1st order peaks related to the smectic layer order decreases with increasing temperature: the 1st order peak increases in intensity, while the 2nd order peak becomes smaller and disappears in the SmA* phase. The $I_{002}/I_{001}$ ratios shown in Figure 3c were corrected by the Lorentz-polarization factor [39] for each peak, as this factor strengthens the observed intensities at low angles. However, the uncorrected $I_{002}/I_{001}$ ratios, although they are smaller than after correction, have qualitatively the same dependence on temperature. The $I_{002}/I_{001}$ values decrease exponentially on heating. This is especially well-visible in the logarithmic scale, where the dependence becomes linear. The fitting parameters of the $I_0 \exp(kT)$ function are $I_0$ = 90(10) and $k$ = −0.0166(4) 1/K. The $I_{002}/I_{001}$ ratio seems insensitive to the glass transition and it also follows the same temperature dependence in the $SmX_A$* and $SmC_A$* phases in the investigated temperature range.

The diffuse maximum at 2θ ≈ 20° is more conveniently analyzed when plotted against the scattering vector $q = 4\pi \sin\theta / \lambda$, where it is described by the Lorentzian peak function, located at $q_0$ and with the full-width at half-height $2/\xi$, where $\xi$ is the correlation length of the short-range positional order [7,40]:

$$I(q) = \frac{A}{1+\xi^2(q-q_0)^2} + Bq + C. \quad (2)$$

The $A$ parameter is the height of the diffuse maximum, $B$ and $C$ are the parameters of the linear background, and the average distance between molecules can be determined as $w = 2\pi/q_0$. There are differences between the results for the supercooled state and those obtained above the melting temperature of a crystal $T_m$ (Figure 3d). In the supercooled and vitrified smectic phase, $w$ = 4.4-4.5 Å and $\xi$ = 11.2-11.9 Å indicate significant nearest-neighbor and next-nearest-neighbor interactions within the smectic layers. Above the melting temperature of a crystal, $w$ = 4.7-4.8 Å and $\xi$ = 4.3-5.0 Å in the smectic phases and $w$ = 4.9 Å and $\xi$ = 4.0 Å in the isotropic liquid, which means only nearest-neighbor interactions. The intermolecular distance matches the molecular width, which is a commonly obtained result [5,41,42]. The matter of discussion is the identification of the smectic phase which undergoes the vitrification for 7HH6, i.e. whether it is the $SmC_A$* phase with only short-range positional order within layers, or the hexatic $SmX_A$* phase ($SmF_A$* or $SmI_A$* – indistinguishable based on the 1D XRD patterns), where the short-range order has a broader range and the bond-orientational order arises [7]. The differential scanning calorimetry results presented in [23] indicate the $SmC_A$* → $SmX_A$* transition at 263.5 K during cooling. This is visible as a small endothermic anomaly with the enthalpy change of −0.34 kJ/mol. The XRD results also show that the correlation length in the 183-263 K range is twice as large as for the $SmC_A$* phase above $T_m$. In similar compounds, which form the $SmC_A$* glass, the correlation length in the vitrified state reaches only 5.2-5.5 Å and does not differ much from the $\xi$ values at higher temperatures [5]. Another indication of the $SmX_A$* phase is an increase of the layer spacing in the supercooled state compared to the $SmC_A$* above $T_m$. The increase of $d$ was reported for other compounds where the transition to the hexatic phase was directly visible in the XRD patterns [10,43-45]. On the other hand, in the XRD patterns of the hexatic phases, the maximum at 2θ ≈ 20° is expected to be sharper (with a spiky peak) [43,46] than the one obtained here for 7HH6. Also, $\xi \approx$ 11-12 Å in the vitrified smectic phase of 7HH6 is a rather low value, since even for the simplest SmA phase, larger intra-layer correlations are reported for some compounds [46,47].

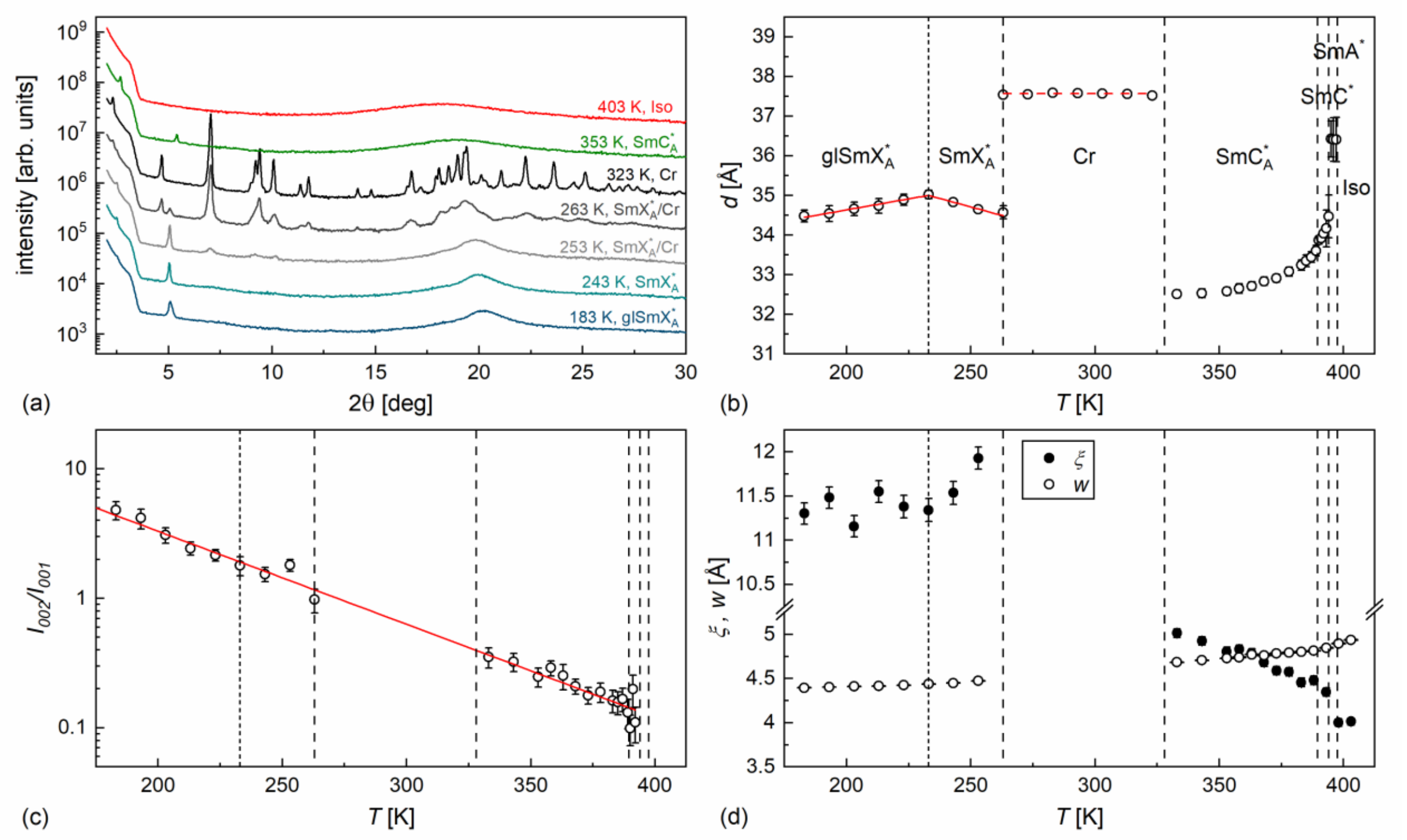

Figure 3. XRD patterns of 7HH6 collected on heating after direct cooling from the $SmC_A$* phase to the vitrified smectic phase with the 6 K/min rate (a) and the temperature dependence of the layer spacing in the smectic and crystal phases (b), ratio of integrated intensities of the (002) and (001) peaks in the smectic phases, corrected by the Lorentz-polarization factor (c) and the correlation length and average intermolecular distance in the smectic and isotropic liquid phases (d).

### *3.2. Tilt angle and selection of molecular models*

The 7HH6 compound forms the orthoconic $SmC_A$* phase, where the tilt angle of molecules $\Theta_O$, determined by the electro-optic method, reaches 44° [16]. The tilt angle decreases with increasing temperature, corresponding to the increasing layer spacing, and attempts are made to find a relationship between these parameters [9-13,48-51]. For the rod-like molecules, the optical tilt angle $\Theta_O$ is equal to the steric tilt angle $\Theta_S = \arccos(d/L)$, where $L$ is the molecular length obtained from molecular modeling or from the layer spacing in the SmA phase with a high orientational order parameter. However, the smectic phases are also formed by molecules with other shapes, like the hockey-stick, zig-zag or C-shape, which gives the molecular core a different orientation than the whole molecule. In calculation of $\Theta_O$, the tilt of the core is important [9-12]. This is why the applied formula is [48]:

$$\Theta_O = \Theta_S + \delta\Theta = \arccos(d/L) + \delta\Theta, \quad (3)$$

where $\delta\Theta$ is the shape parameter, equal to an angle between the direction of the whole molecule and direction of the aromatic core. Several molecular conformations were tested in [48] to check which one enables the best reproduction of $\Theta_O$ from the layer spacing obtained by XRD. In this paper, another approach is applied: the layer spacing from XRD is plotted against the tilt angle from electro-optic measurements (Figure 4a) and fitted with the formula:

$$d = L\cos(\Theta_O - \delta\Theta), \quad (4)$$

which enables determination of $L$ and $\delta\Theta$. The knowledge of these parameters is further used to infer the most likely conformations of the molecule. For 7HH6 the fitting results are $L$ = 34.2(2) Å and $\delta\Theta$ = 25.1(8)° in the SmC$_A$* phase, and $L$ = 35.6(4) Å and $\delta\Theta$ = 19(2)° in the SmC* phase. The increase of the molecular length and decrease of the shape parameter at the SmC$_A$* → SmC* transition indicate that molecules take a more extended shape in SmC*. The $L$ values are smaller than the layer spacing 36.4(6) Å in SmA*, indicating that in the SmA* phase, molecules have a more extended conformation. In both SmC$_A$* and SmC* phases, the difference between the overall tilt of molecules and the tilt of the aromatic core is significant, as it equals more than half of $\Theta_O$. The fitting with an assumption of the rod-like molecule with $\delta\Theta$ = 0, performed as a test, gives results in a considerable disagreement with the experimental data. In some papers, the discrepancy between $\Theta_O$ and $\Theta_S$ is explained by the dimer formation, where the length of a dimer exceeds the length of a single molecule [13,49,50]. However, for 7HH6 the fitting $L$ values indicate that the main building block is a single molecule, not a dimer.

The main purpose is to find the molecular models with the $L$ and $\delta\Theta$ values close to the fitting results of the (4) formula. The DFT results show at least two such models (Figure 3b). The definitions of $L$ and $\delta\Theta$ and the molecular conformations included in this analysis are the same as in [48]. The length of a molecule $L$ is equal to $L_0 + L_{CF}$, where $L_0$ is the length of the $\overrightarrow{L_0}$ vector from the terminal C atom in the chiral chain to this terminal F atom in the achiral chain which lies approximately in the same plane as the preceding C atoms, and $L_{CF}$ = 3.22 Å is the non-bonding distance between C and F [51]. The orientation of the aromatic core is described by the $\vec{A}$ vector from the C atom in the COO group, located near the C* atom, to the O atom adjacent to the benzene ring. The shape parameter $\delta\Theta$ is an angle formed by the $\overrightarrow{L_0}$ and $\vec{A}$ vectors (Figure 5a). The torsional angles selected for the optimization are the $\varphi_1$ angle formed by the C($H_2$)-C($H_2$)-C*($HCH_3$)-O atoms and describing the orientation of the chiral terminal chain, as well as the $\varphi_2$ and $\varphi_3$ angles formed by the C($H_2$)-O-C($H_2$)-C($F_2$) and O-C($H_2$)-C($F_2$)-C($F_2$) atoms, describing the orientation of the fluorinated part of the achiral terminal chain (Figure 5b).

Using the three local minima in the conformation energy for each torsional angle (Figure 5b), 27 final molecular models were obtained, and for each model, the tilt angle from the (3) formula was determined (Table 1). Eight models are giving at 333 K the $\Theta_O$ angle which agrees, within the ±2° error, with $\Theta_O$ = 43.8° measured electro-optically at the same temperature in Ref. [16]. Among them, only models 3 and 19 reproduce correctly the $\Theta_O(T)$ dependence in almost the whole considered temperature range, except the proximity of the SmC*/SmA* transition. The model 19 has a lower conformational energy, on the other hand, the model 3 was considered more probable in [48], as it led to correct $\Theta_O$ values for both compounds investigated therein. Different $L$ and $\delta\Theta$ obtained for both phases in Figure 4a indicate that the molecular conformations in the $SmC_A$* and SmC* phase slightly differ. Interestingly, $L_3 < L_{19}$ and $\delta\Theta_3 > \delta\Theta_{19}$, which agrees with the relationships between the $L$ and $\delta\Theta$ values for the $SmC_A$* and SmC* phases. Models 3 and 19 correspond to the same local energy minimum for the $\varphi_2$ angle. The changes in the $\varphi_1$ and $\varphi_3$ angles, necessary for transition between these models, are related to the energy barriers of 16 kJ/mol and 18 kJ/mol. Thus, it is not certain whether the $SmC_A$*/SmC* transition is related to the change between these two particular conformation, because the enthalpy change at this transition is negligible (>1 kJ/mol [2,23]). The more likely explanation is that the $SmC_A$*/SmC* transition does not involve the conformational changes in all molecules but rather the changes in the fractions of molecules taking particular conformations.

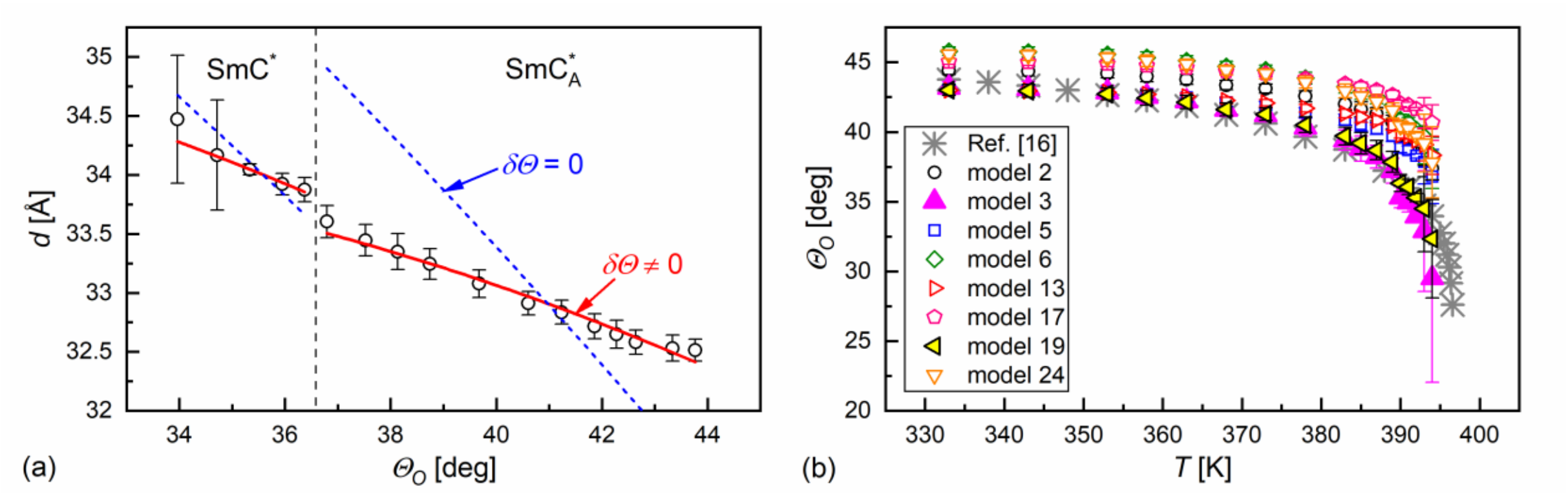

Figure 4. Smectic layer spacing determined from the XRD patterns (this work) vs. tilt angle measured by the electro-optic method at corresponding temperatures (values from Ref. [16]) with the fitting results of the (4) formula (a) and the tilt angle calculated from the (3) formula for various molecular models compared with the $\Theta_O$ values from Ref. [16] as a function of temperature (b).

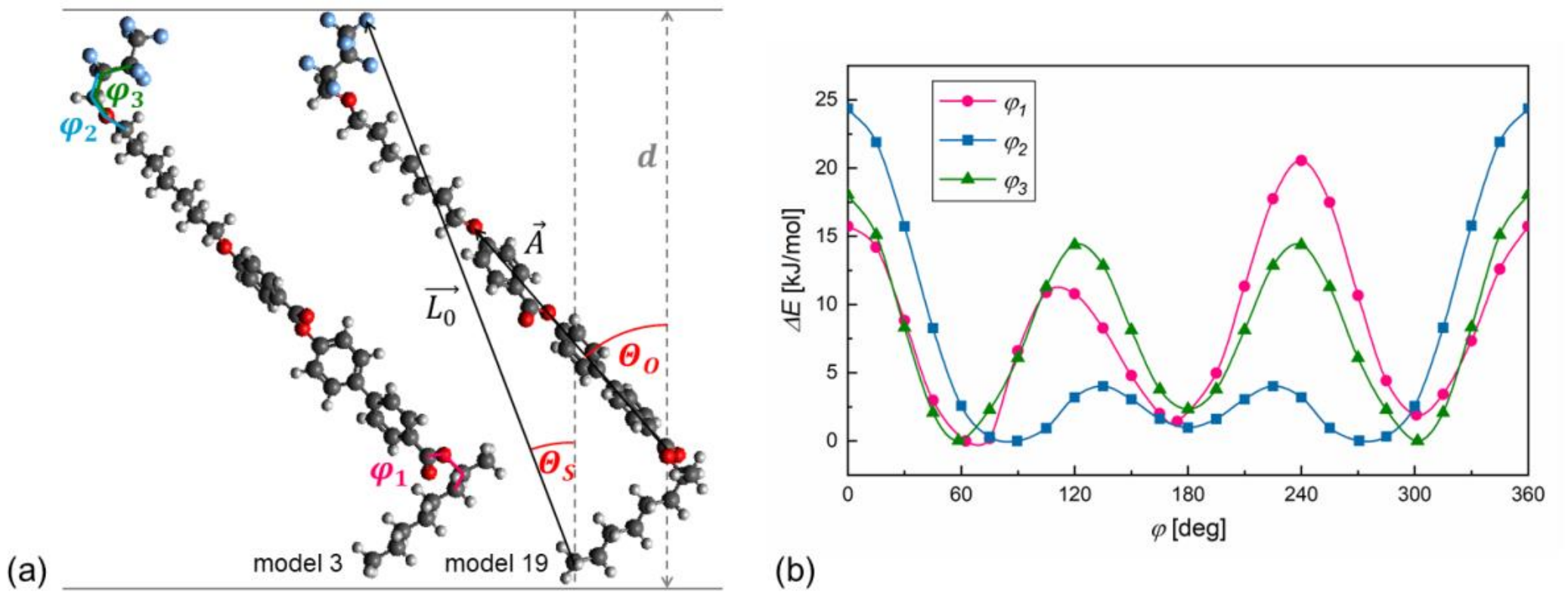

Figure 5. Molecular models 3 and 19 of 7HH6 with indicated torsional angles and vectors used in calculations of the tilt angle (a) and energy scans for torsional angles $\varphi_1$, $\varphi_2$, $\varphi_3$ (b).

Table 1. DFT results (def2TZVPP basis set, B3LYP-D3(BJ) functional) for 27 conformations of the 7HH6 molecule, corresponding to local minima in energy presented in Figure 5b. For each conformation, the tilt angle was obtained according to the (3) formula, using the smectic layer spacing 32.5(1) Å in 333 K. The **bold** font indicates the models that agree, within the ±2° error, with the experimental tilt angle of 43.8° [16] in 333 K.

| model | $\Delta E$ [kJ/mol] | $\varphi_1$ [deg] | $\varphi_2$ [deg] | $\varphi_3$ [deg] | $L$ [Å] | $\delta\Theta$ [deg] | $\Theta_O$ [deg] |
|---|---|---|---|---|---|---|---|
| 1 | 0.02 | 62.3 | 79.0 | 50.2 | 35.9 | 21.4 | 46.4(4) |
| **2** | **2.81** | **62.3** | **89.5** | **176.2** | **37.2** | **15.4** | **44.4(3)** |
| **3** | **4.17** | **62.3** | **109.9** | **294.7** | **34.7** | **22.6** | **43.2(5)** |
| 4 | 1.48 | 62.4 | 181.3 | 58.5 | 36.6 | 12.9 | 35.1(4) |
| **5** | **3.79** | **62.4** | **180.0** | **180.1** | **38.0** | **11.8** | **43.1(3)** |
| **6** | **1.46** | **62.4** | **178.7** | **301.5** | **37.3** | **15.8** | **45.8(4)** |
| 7 | 4.18 | 62.2 | 250.1 | 65.5 | 32.4 | 4.2 | - |
| 8 | 2.83 | 62.4 | 270.6 | 184.2 | 36.6 | 9.8 | 37.0(4) |
| 9 | 0 | 62.2 | 281.0 | 309.9 | 33.8 | 4.9 | 20.6(6) |
| 10 | 1.45 | 174.4 | 78.9 | 50.5 | 40.6 | 13.6 | 50.5(3) |
| 11 | 4.24 | 174.3 | 90.2 | 176.8 | 42.5 | 10.1 | 50.3(2) |
| 12 | 5.64 | 174.4 | 110.1 | 294.5 | 39.4 | 15.1 | 49.5(3) |
| **13** | **2.87** | **174.2** | **181.5** | **58.4** | **41.2** | **5.1** | **43.0(3)** |
| 14 | 5.21 | 174.2 | 180.5 | 180.2 | 43.4 | 6.2 | 47.6(2) |
| 15 | 2.90 | 174.2 | 178.6 | 301.6 | 41.8 | 11.2 | 50.0(2) |
| 16 | 5.61 | 174.3 | 250.0 | 65.4 | 38.2 | 2.0 | 33.7(3) |
| **17** | **4.25** | **174.4** | **270.5** | **184.1** | **42.2** | **5.5** | **45.0(2)** |
| 18 | 1.43 | 174.4 | 280.8 | 309.8 | 39.6 | 0.5 | 35.4(3) |
| **19** | **1.89** | **300.9** | **78.9** | **50.0** | **35.2** | **20.3** | **43.0(4)** |
| 20 | 4.72 | 300.9 | 89.8 | 176.3 | 36.6 | 13.2 | 40.5(4) |
| 21 | 6.08 | 300.8 | 109.6 | 294.5 | 33.9 | 21.0 | 37.2(6) |
| 22 | 3.36 | 300.8 | 181.1 | 58.4 | 35.8 | 13.4 | 38.1(4) |
| 23 | 5.70 | 300.9 | 180.3 | 180.1 | 37.5 | 10.9 | 40.8(3) |
| **24** | **3.37** | **300.7** | **178.9** | **301.5** | **36.6** | **18.4** | **45.6(4)** |
| 25 | 6.10 | 300.8 | 250.5 | 65.5 | 32.1 | 5.6 | - |
| 26 | 4.73 | 300.9 | 269.5 | 183.1 | 35.9 | 6.9 | 32.2(4) |
| 27 | 1.92 | 300.8 | 281.1 | 309.9 | 33.6 | 7.6 | 22.4(7) |

*3.3. Vibrational modes investigated by Fourier-transform infra-red spectroscopy*

The theoretical IR spectra were calculated for the molecular models 3 and 19 selected in the previous section (Figure 6a). Each model enables the band assignment of the experimental IR spectra collected in the $SmC_A$* phase at 333 K (Table 2). The bands in the 513-661 $cm^{-1}$ range correspond mainly to the scissoring and rocking vibrations in the terminal chains, as well as in-plane and out-of-plane deformations of the aromatic rings, 692-771 $cm^{-1}$ to the $CH_2$ rocking and out-of-plane deformations of the aromatic rings, 815-912 $cm^{-1}$ to the rocking and stretching vibrations in the aromatic chains, and out-of-plane deformations of the aromatic rings, 961-1070 $cm^{-1}$ to the C-C, C-O stretching and in-plane deformations of the aromatic rings, 1111-1275 $cm^{-1}$ to the $CH_2$ rocking and wagging, C-C, C-O, C-F stretching, and in-plane deformations of the aromatic rings, 1314-1422 $cm^{-1}$ to the $CH_2$ twisting and wagging, and in-plane deformations of the aromatic rings, and 1510-1607 $cm^{-1}$ to the $CH_2$ scissoring and in-plane deformations of the aromatic rings. The band at 1711 $cm^{-1}$ is attributed to the C=O stretching in the COO(h) group next to the chiral center, and the band at 1733 $cm^{-1}$ to the C=O stretching in the COO(e) spacer within the aromatic core. The bands in 2862-2981 $cm^{-1}$ and 3042-3074 $cm^{-1}$ correspond to the C-H stretching in the terminal chains and in the aromatic rings, respectively. The overall scaling factor between the calculated and experimental wavenumbers, obtained based on the DFT results for both models (Figure 6b), equals 0.969(2). Since the scaling may vary between different frequency ranges [52], the scaling factor was also obtained separately for experimental values >1000 $cm^{-1}$, 1000-2000 $cm^{-1}$ and >2000 $cm^{-1}$, and is equal to 0.977(3), 0.986(2) and 0.955(2), respectively. Thus, the most significant overestimation in the calculated wavenumbers occurs for the bands originating from the C-H stretching. The corresponding results for the IR spectrum of the pristine sample at 293 K are presented in Electronic Supplementary Materials together with atomic coordinates and IR spectra calculated for the models 3 and 19. The scaling factors for the pristine sample at 293 K agree within uncertainties with these for the $SmC_A$* phase at 333 K.

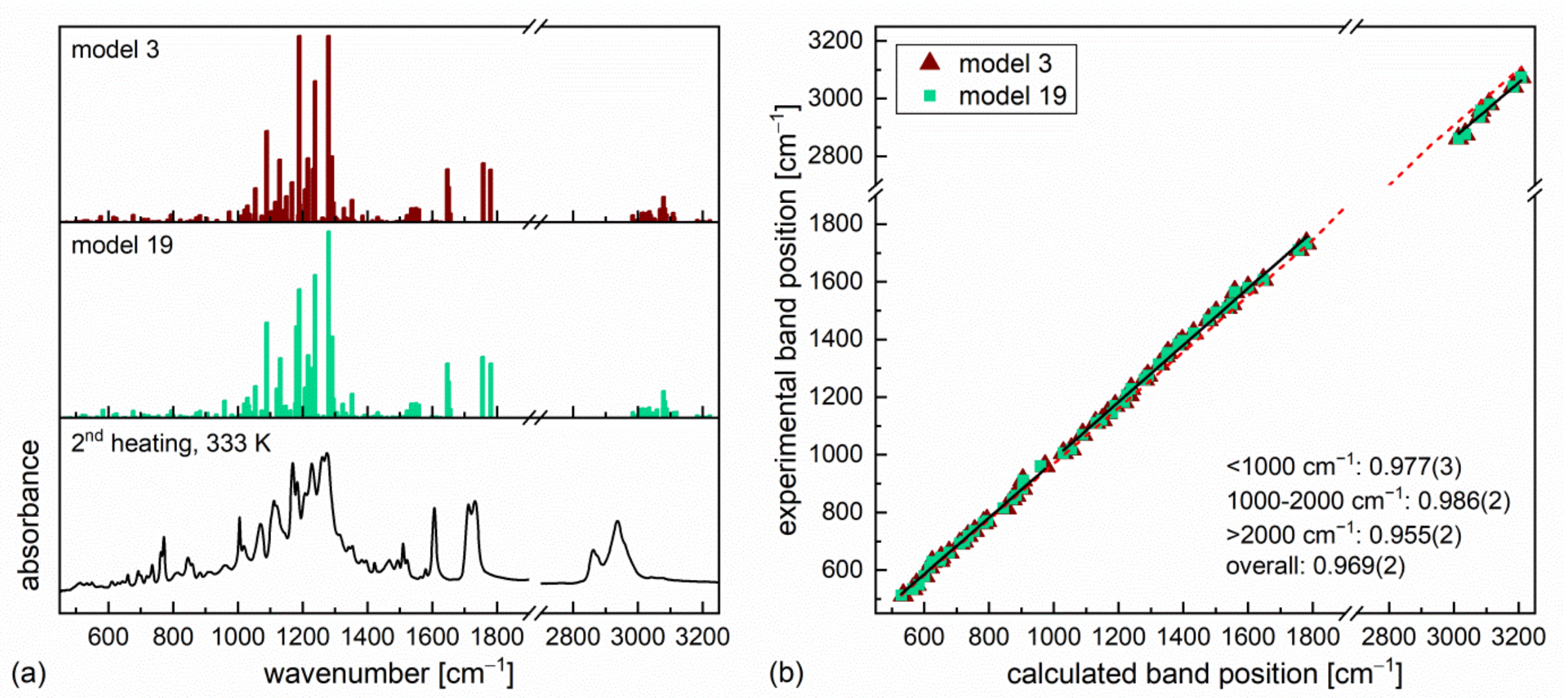

Figure 6. The IR spectrum of 7HH6 collected upon the 2nd heating at 333 K compared to the unscaled calculated spectra for the molecular models 3 and 19 (DFT, def2TZVPP basis set, B3LYP-D3(BJ) functional) (a) and scaling between the experimental and calculated band positions, $\tilde{\nu}_{exp} = a\tilde{\nu}_{calc}$, where $a$ is the scaling factor (b).

Table 2. The band assignment of the experimental IR spectra of 7HH6 measured upon the 2nd heating at 333 K, based on the DFT calculations (def2TZVPP basis set, B3LYP-D3(BJ) functional). Experimental and calculated wavenumbers are given in cm$^{-1}$. The intra-molecular vibrations are denoted as follows: β – in-plane deformation, γ – out-of-plane deformation, δ – scissoring, ν – stretching, ρ – rocking, τ – twisting, ω – wagging.

| experimental | model 3 | model 19 | interpretation |
|---|---|---|---|
| 513 | 535.8 | 529.5 | δ$CF_2$(a), δCOC(b,c), δCCC(c), γPh(f,g), δOCC(b,c)$^{model3}$, $β_{asym}$Ph(d)$^{model3}$, γPh(d)$^{model19}$ |
| 534 | 565.4 | 565.1 | γPh(f,g) |
| 550 | 576.2 | 584.0 | δ$CF_2$(a), δCCO(a,b) |
| 579 | 602.4 | 600.5 | ρ$CF_2$(a), ρ$CH_2$(b) |
| 610 | 616.6 | 617.3 | δ$CF_2$(a), ρ$CH_2$(b) |
| 629 | 624.7 | 624.7 | $β_{asym}$Ph(d,f,g) |
| 633 | 648.7 | 648.8 | $β_{asym}$Ph(d,g), γPh(f) |
| 644 | 651.9 | 651.9 | $β_{asym}$Ph(d,f,g) |
| 661 | 677.3 | 677.1 | $β_{asym}$Ph(d,g), γPh(f) |
| 692 | 709.9 | 710.0 | γPh(d) |
| 700 | 721.7 | 724.4 | γPh(f,g) |
| 719 | 735.6 | 735.9 | ρ$CH_2$(j) |
| 736 | 755.7 | 754.4 | ρ$CH_2$(c) |
| 763 | 783.5 | 783.4 | γPh(d) |
| 771 | 793.8 | 795.2 | γPh(d,f,g), ρ$CH_2$(j) |
| 815 | 853.6 | 844.2 | τ$CH_2$(j)$^{model3}$, γPh(f)$^{model19}$ |
| 846 | 872.6 | 872.7 | γPh(d) |
| 859 | 883.1 | 883.6 | γPh(f,g) |
| 883 | 904.1 | 904.5 | νCC(a,b)$^{model3}$, ρ$CH_2$(b)$^{model3}$, γPh(f,g)$^{model19}$ |
| 912 | 904.2 | 906.0 | γPh(f,g)$^{model3}$, νCC(a,b)$^{model19}$, ρ$CH_2$(b)$^{model19}$ |
| 961 | 973.7 | 959.1 | $\nu_{sym}$COC(b,c), $\nu_{sym}$CCC(c) |
| 1006 | 1029.8 | 1029.7 | $\beta_{asym}$Ph(d) |
| 1019 | 1054.4 | 1054.6 | $\nu_{asym}$CCC(c), $\nu$CO(c), $\beta_{asym}$Ph(d,f,g) |
| 1070 | 1089.4 | 1089.3 | $\beta_{asym}$Ph(d,f), $\nu$CO(e) |
| 1111 | 1129.3 | 1131.6 | $\beta_{asym}$Ph(g), $\nu_{asym}$COC(h,i) |
| 1121 | 1149.8 | 1147.9 | $\nu_{asym}CF_2$(a), $\nu_{sym}$CCC(c), $\beta_{sym}$Ph(d), $\nu_{asym}$COC(b,c)$^{model3}$, $\rho CH_2$(b)$^{model19}$ |
| 1145 | 1167.1 | 1181.5 | $\nu_{asym}CF_2$(a), $\rho CH_2$(b), $\nu_{asym}$COC(b,c)$^{model19}$ |
| 1170 | 1189.5 | 1189.5 | $\beta_{sym}$Ph(d,f) |
| 1183 | 1216.6 | 1217.8 | $\omega CF_2$(a), $\omega CH_2$(b) |
| 1207 | 1233.1 | 1225.7 | $\nu_{sym}CF_2$(a), $\nu$CC(a), $\delta$CCC(a) |
| 1229 | 1238.5 | 1238.5 | $\beta_{sym}$Ph(d,g), $\beta_{asym}$Ph(f), $\nu_{asym}$COC(e,f) |
| 1261 | 1280.4 | 1280.5 | $\beta_{asym}$Ph(d,f), $\nu_{asym}$CCO(d,e) |
| 1275 | 1290.7 | 1291.7 | $\beta_{asym}$Ph(g), $\nu_{asym}$CCO(g,h), $\delta$OC*H(h,i) |
| 1314 | 1327.7 | 1323.9 | $\omega CF_2$(a), $\tau CH_2$(b,c) |
| 1342 | 1345.2 | 1349.3 | $\nu_{sym}$CCC(a), $\tau CH_2$(b,c), $\omega CH_2$(c), $\beta_{asym}$Ph(d) |
| 1354 | 1352.6 | 1353.0 | $\tau CH_2$(b), $\omega CH_2$(c), $\beta_{asym}$Ph(d), $\tau CH_2$(c)$^{model19}$ |
| 1383 | 1385.1 | 1385.1 | $\omega CH_2$(i,j) |
| 1396 | 1397.3 | 1397.5 | $\omega CH_2$(b,c) |
| 1422 | 1432.0 | 1432.3 | $\omega CH_2$(c), $\beta_{asym}$Ph(d) |
| 1468 | 1477.9 | 1478.2 | $\delta CH_2$(i,j)$^{model3}$, $\delta CH_2$(b,c)$^{model19}$, $\omega CH_2$(c)$^{model19}$ |
| 1494 | 1501.1 | 1501.1 | $\delta CH_3$(j) |
| 1510 | 1536.4 | 1536.5 | $\beta_{asym}$Ph(f,g) |
| 1523 | 1550.0 | 1550.1 | $\beta_{asym}$Ph(d) |
| 1565 | 1559.3 | 1559.3 | $\beta_{asym}$Ph(f,g) |
| 1580 | 1600.3 | 1600.3 | $\beta_{asym}$Ph(f,g) |
| 1607 | 1647.5 | 1647.5 | $\beta_{sym}$Ph(d,f,g) |
| 1711 | 1758.2 | 1756.1 | $\nu$C=O(h) |
| 1733 | 1780.8 | 1781.2 | $\nu$C=O(e) |
| 2862 | 3014.7 | 3014.4 | $\nu_{sym}CH_2$(c), $\nu_{asym}CH_2$(j) |
| 2876 | 3035.5 | 3034.9 | $\nu_{sym}CH_2$(c), $\nu_{sym}CH_2$(b)$^{model3}$, $\nu_{asym}CH_2$(c)$^{model19}$ |
| 2937 | 3079.5 | 3079.4 | $\nu_{asym}CH_2$(c) |
| 2959 | 3083.7 | 3083.3 | $\nu_{asym}CH_2$(j) |
| 2981 | 3109.1 | 3110.0 | $\nu CH_3$(i) |
| 3042 | 3183.6 | 3183.6 | $\nu_{asym}$CH(f,g) |
| 3074 | 3208.1 | 3207.6 | $\nu_{sym}$CH(f,g) |

Next, the temperature dependence of the experimental IR spectra is investigated. At the room temperature, the pristine 7HH6 sample is in the crystal phase. After the 1$^{st}$ heating to the isotropic liquid and cooling down with the 3 K/min rate to 173 K, the vitrified SmX$_A$* phase is obtained. Upon the 2$^{nd}$ heating, the cold crystallization and subsequent melting of the crystal phase are observed (Figure 7a). The bands from the C=O stretching are often sensitive to the phase transitions [21,53,54]. They are also well separated from other absorption bands, therefore, they could have been deconvolved by fitting the peak functions (Figure 7b-d). They are denoted as I, II, III, and IV in an order of an increasing wavenumber. There are four $\nu$C=O bands in the crystal phase, three in the SmX$_A$* phase and the SmX$_A$* glass, and two in the SmC$_A$* and Iso phases. The SmC* and SmA* phases have a narrow temperature range [23] and the IR spectrum collected at 393 K, which belongs to SmC* or SmA*, also shows two $\nu$C=O bands. Bands I and IV are observed in all phases, band III appears in the crystal and SmX$_A$* phases, while band II is observed only in the crystal phase (Figure 8). The appearance of additional νC=O bands is usually caused by inter- or intramolecular interactions, which have various effects: the hydrogen bonding shifts the νC=O band towards lower wavenumbers (redshift) [55-57]. In comparison, the proximity of the F atom shifts the νC=O band towards higher wavenumbers (blueshift) [58]. As 7HH6 is not fluorosubstituted in the molecular core, the latter effect can be caused only by proximity to a neighbor molecule's partially fluorinated terminal chain. The interpretation of bands I-IV is presented in Table 3. The bands I and IV are attributed to the C=O stretching in the COO groups (h) and (e), respectively, which are not close to the F atom. Both bands I and IV shift to higher wavenumbers upon heating above the melting temperature, which may be the effect of the breaking of very weak intermolecular hydrogen bonds involving the COO groups, e.g., of the C-H…O type. A lower wavenumber of band I below the melting temperature indicates stronger hydrogen bonding involving the COO (h) group in the $SmX_A$* and Cr phases than in $SmC_A$*. The decrease of the area and increase of band IV's peak position and the appearance of band III in the $SmX_A$* phase suggest that a small fraction of molecules is not hydrogen-bonded. In contrast, most molecules form hydrogen bonds (stronger than in the $SmC_A$* phase) involving the COO (e) group. The possible scenario is that the hydrogen bonds exist between the molecules forming clusters with a local hexagonal ordering, making them larger and more stable than in the $SmC_A$* phase, and the molecules at the borders between clusters are not hydrogen-bonded. In the crystal phase, four νC=O bands indicate that the antisymmetric unit of the crystal structure probably contains two molecules with different surrounding, which are consequently involved in different intermolecular interactions. Thus, the presence of band II can be caused by the proximity between the COO (h) group from one molecule and the F atom from a neighbor molecule or by the lack of hydrogen bonding, and band III corresponds to the hydrogen-bonded COO (e) group.

Table 3. Interpretation of the experimental C=O stretching bands registered for 7HH6.

| phase | I | II | III | IV |
|---|---|---|---|---|
| gl$SmX_A$*/$SmX_A$* | COO(h)<br>H-bonded | - | COO(e)<br>weakly H-bonded | COO(e)<br>not H-bonded |
| Cr | COO(h)<br>H-bonded | COO(h)<br>not H-bonded<br>or close to F | COO(e)<br>H-bonded | COO(e)<br>not H-bonded |
| $SmC_A$*/SmC*/SmA* | COO(h)<br>weakly H-bonded | - | - | COO(e)<br>weakly H-bonded |
| Iso | COO(h)<br>not H-bonded | - | - | COO(e)<br>not H-bonded |

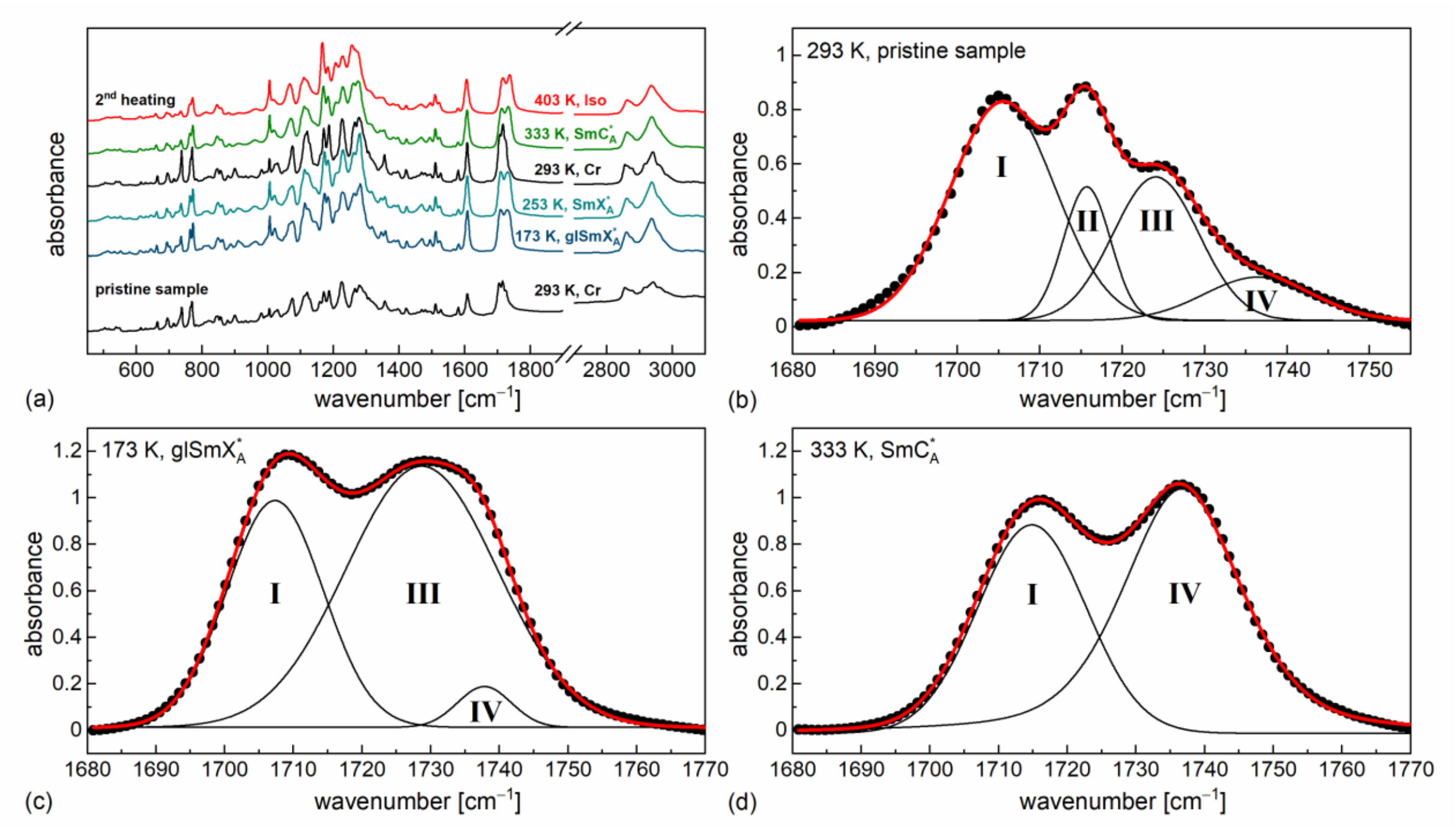

Figure 7. IR spectra of 7HH6 collected at 293 K for a pristine sample and upon heating from the vitrified $SmX_A$* phase (a), and exemplary fits of pseudo-Voigt peak functions to the absorption bands related to the C=O stretching for the crystal phase at 293 K (a), vitrified $SmX_A$* phase at 173 K (c) and $SmC_A$* phase at 333 K (d). The results in (b) and (c,d) were obtained for the 1st and 2nd heating runs. The baseline was manually subtracted from (b-d) before fitting.

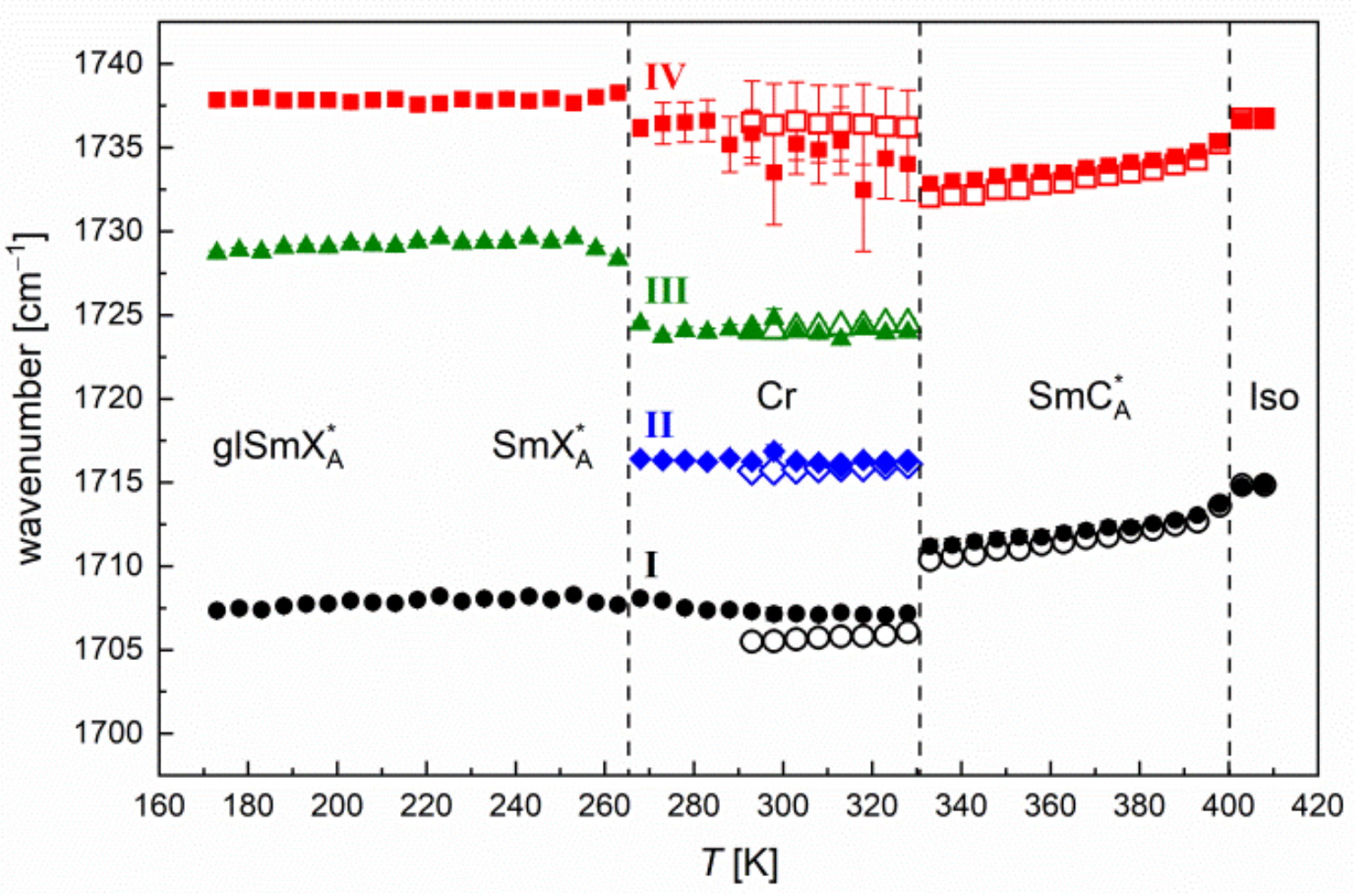

Figure 8. Wavenumbers of the νC=O bands vs. temperature, obtained from the IR spectra of 7HH6 collected upon the 1st heating of the pristine sample (open symbols) and 2nd heating of the vitrified sample (solid symbols). The SmC* and SmA* phases are not indicated in a narrow temperature range between $SmC_A$* and Iso.

The positions of the νC=O bands are sensitive to the cold crystallization, melting and transition to the isotropic liquid of 7HH6, while they do not show a clear change at $T_g$. Also other parameters of the νC=O bands: area, half-width and height, when plotted as a function of temperature, are too scattered to reveal the glass transition. The exceptions are the height and area of the weak peak IV, which both show a minimum in the glass transition temperature region (Figure S2).

Better insight into the influence of phase transitions on the intra-molecular vibrations can be obtained by the k-means cluster analysis [27,59-62]. This method enables the segregation of data – in this case, the IR spectra collected at different temperatures – into mentioned clusters. The data within each cluster are maximally similar and differ maximally between the clusters [27,59-61]. The analysis can be done for the spectra in the whole 450-4000 cm$^{-1}$ range. However, as presented in the papers [61,62], performing it separately for selected spectral ranges is more informative. For the IR spectra registered as a function of temperature, one can infer which intra-molecular vibrations are more affected by the phase transition [61]. Thus, each IR spectrum of 7HH6 was divided into 10 parts (Figure 9) and a cluster analysis was performed on each one independently. The number of clusters optimal for a given dataset can be calculated using elbow method [59,60], based on the summed square error plotted vs. the number of clusters. The optimal number of clusters is presumed to be at the inflection point. The analysis by the elbow method for the IR spectra of 7HH6 in the 1650-1850 cm$^{-1}$ range suggests that the optimal number of clusters is 3 (Figure S3), and it was applied to the results in all spectral ranges (Figure 9a). Cluster 1 contains the IR spectra collected in the 173-263 K or 173-268 K temperature ranges ($SmX_A$* and its glass), cluster 2 – spectra from 268-328 K or 273-328 K (crystal phase), and cluster 3 – spectra from 333-408 K ($SmC_A$*-isotropic liquid phases). The temperature of cold crystallization determined by the cluster analysis is ~268 K, which is higher than 253 K observed in the XRD patterns. It is caused by the fact that the XRD measurement is slower than the IR spectroscopy one, therefore, the sample has time to crystallize at a lower temperature. The melting temperature of a crystal is in agreement with the XRD results.

In the next step, the IR spectra from 268-328 K were excluded, and the cluster analysis was performed separately for the 173-263 K and 333-408 K ranges. For each temperature range, two clusters were assumed, corresponding to gl$SmX_A$*, $SmX_A$* for 173-263 K and $SmC_A$*, Iso for 333-408 K. The 2-cluster analysis for 173-263 K (Figure 9b) shows that the absorption bands from the 1090-1300 cm$^{-1}$ and 1650-3200 cm$^{-1}$ ranges are divided into two clusters in a random manner. This indicates that the experimental data's noise is stronger than any changes originating from the glass transition. The bands from the 450-675 cm$^{-1}$ and 1300-1550 cm$^{-1}$ ranges indicate $T_g$ ~ 218 K, while bands from the 675-1090 cm$^{-1}$ and 1550-1650 cm$^{-1}$ ranges indicate $T_g$ ~ 228 K. The latter is closer to $T_g$ = 229 K determined from the dielectric spectra [23] and $T_g$ = 233 K obtained in this work from the XRD patterns.

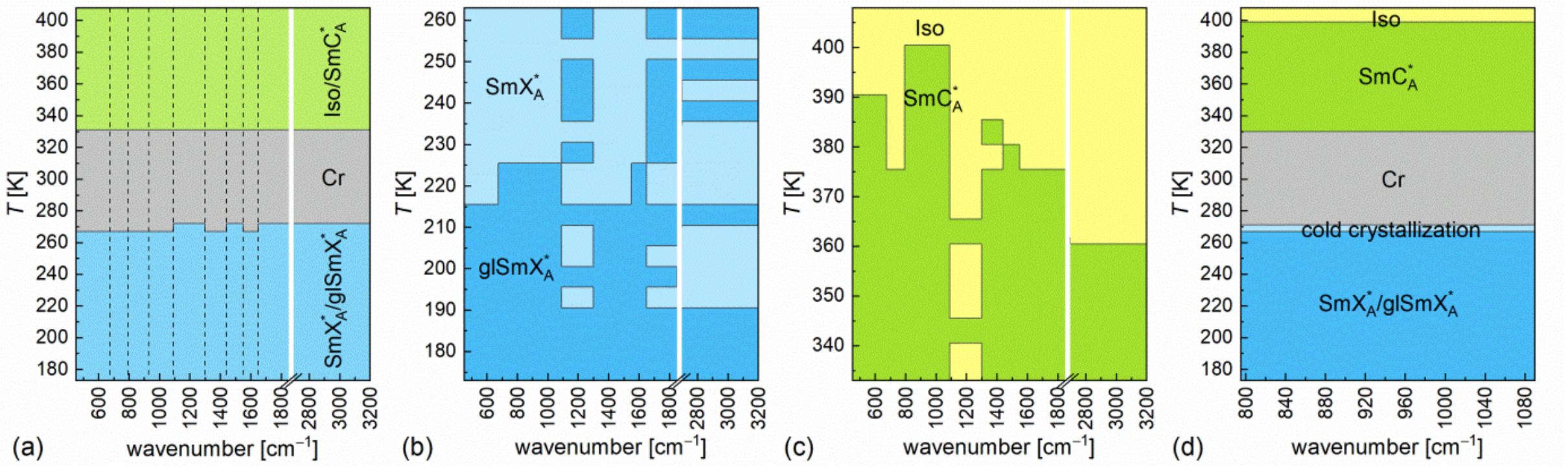

Figure 9. The phase sequence of the vitrified 7HH6 sample on heating determined by the k-means cluster analysis for various ranges in the IR spectra: 3-cluster analysis in the 173-403 K range (a), 2-cluster analysis in 173-263 K (b) and 333-408 K (c), and 5-cluster analysis in 173-403 K for the spectral range 795-1090 $cm^{-1}$. The vertical dashed lines in the panel (a) indicate the separate spectral ranges.

The 2-cluster analysis for 333-408 K (Figure 9c) gives surprising results because only the absorption bands from the 795-1090 $cm^{-1}$ range are correctly divided into the $SmC_A$* and Iso phases. Even for the νC=O bands, the border between clusters is at 378 K, much lower than the transition temperature to the isotropic liquid ~400 K determined from the νC=O band positions (Figure 8). It shows that a cluster analysis should not be used as the only way of data analysis because it can be inappropriate in some cases. For the νC=O bands, the indicator of the smectic – liquid transition is the change in the temperature dependence: increasing band position in $SmC_A$* and constant in Iso, not the distances between values for which the algorithm is sensitive.

The bands in the 795-1090 $cm^{-1}$ range proved to be the most effective when used in the cluster analysis, as they gave satisfactory results both for 173-263 K and 333-408 K. They originate from the vibrations in the achiral chain (C-C, C-O stretching, $CH_2$ rocking) and the aromatic core (in-plane and out-of-plane deformations). Only one band in this range may originate from the $CH_2$ twisting in the chiral chain, according to molecular model 3 (Table 2). In the last step, the 5-cluster analysis was performed for the 795-1090 $cm^{-1}$ range for all temperatures, to check if the phase sequence can be reproduced (Figure 9d). However, while the IR spectra are correctly attributed to the Cr, $SmC_A$* and Iso phases, the $SmX_A$* and its glass are within the same cluster. The fifth cluster contains only the IR spectrum from 268 K, which corresponds to the cold crystallization and was attributed by the 3-cluster analysis either to the $SmX_A$* or Cr phase (Figure 9a). It shows that in the investigation of the glass transition, it is more convenient to consider only the temperature range of the supercooled state because stronger effects related to the cold crystallization may disturb the cluster analysis.

## 4. Conclusions

The 7HH6 compound, forming four smectic phases: SmA*, SmC*, $SmC_A$*, and hexatic $SmX_A$*, was investigated by X-ray diffraction and IR spectroscopy:

- The transitions between the smectic phases and the glass transition of the $SmX_A$* phase are detectable in the temperature dependence of the smectic layer spacing. The ratio of the intensity of the 2$^{nd}$ and 1$^{st}$ order peaks, describing the smectic layer order, decreases logarithmically on heating, also below the glass transition temperature.
- The correlation length of the positional short-range order is unusually low in the $SmX_A$* phase, as it extends only to the next-nearest neighbors. However, it is larger than in the $SmC_A$* phase, where it corresponds only to the distance between the nearest-neighbors.
- By analysis of the smectic layer spacing in the $SmC_A$* and SmC* phases as a function of the tilt angle, it is proven that the building block of the smectic layer is a single molecule (not a dimer) with the aromatic core tilted by ca. 20-25° in respect to the whole molecule. Thus, half of the high tilt angle exhibited by 7HH6 in the $SmC_A$* phase originates from the molecular shape.
- The molecular arrangement in the $SmX_A$* phase differs, as the smectic layer spacing is larger than in $SmC_A$*. Another sign of the structural changes affecting the intra-molecular interactions in $SmX_A$* is appearance of the third C=O stretching band, absent in $SmC_A$*, where two C=O stretching bands are observed, each originating from another COO group.
- The experimental positions of the C=O stretching bands, obtained as a function of temperature, suggest that the COO groups in some phases are involved in weak hydrogen bonds.
- The k-means cluster analysis enables the determination of the glass transition temperature only for some spectral ranges. The 795-1090 cm$^{-1}$ range, containing mainly the absorption bands originating from the vibrations in the achiral chain and aromatic core, is the most efficient for detecting the phase transitions by a cluster analysis.

It is planned to prepare the mixture of 3F7HPhH6 with MHPOBC, which also shows the hexatic smectic phase [63]. It will enable checking whether these compounds exhibit the same type of the $SmX_A$* phase ($SmF_A$* or $SmI_A$*).

**Authors' contributions:**

A. Deptuch – conceptualization, investigation (XRD, DFT), formal analysis (XRD, FT-IR), writing – original draft.

N. Górska – investigation (FT-IR), writing – review and editing.

S. Baran – investigation (XRD), writing – review and editing.

M. Urbańska – resources (sample's synthesis), writing – review and editing.

**Declaration statement:** The authors declare no conflict of interest.

**Acknowledgements.** We thank Assoc. Prof. Ewa Juszyńska-Gałązka from Institute of Nuclear Physics Polish Academy of Sciences for the discussion regarding the data analysis and presentation, and Assoc. Prof. Wojciech Zając from Institute of Nuclear Physics Polish Academy of Sciences for help with the DFT calculations. We gratefully acknowledge Poland's high-performance Infrastructure PLGrid Academic Computer Centre Cyfronet AGH for providing computer facilities and support within the computational grant. Empyrean 2 (PANalytical) diffractometer with Cryostream 700 Plus (Oxford Cryosystems) temperature attachment were purchased thanks to European Regional Development Fund Operational Program Infrastructure and Environment (contract no. POIS.13.01.00-00-062/08). Bruker VERTEX 70v FT-IR spectrometer with Advanced Research System DE-202A cryostat and ARS-2HW compressor were purchased thanks to the European Regional Development Fund in the framework of the Innovative Economy Operational Program (contract no. POIG.02.01.00-12-023/08). This research was supported in part by the Excellence Initiative – Research University Program at the Jagiellonian University in Kraków.

**Structural and dynamical investigation of glassforming smectogen by X-ray diffraction and infra-red spectroscopy aided by density functional theory calculations**

Aleksandra Deptuch[1,*], Natalia Górska[2], Stanisław Baran[3], Magdalena Urbańska[4]

[1] Institute of Nuclear Physics Polish Academy of Sciences, Radzikowskiego 152, PL-31342 Kraków, Poland

[2] Faculty of Chemistry, Jagiellonian University, Gronostajowa 2, PL-30387 Kraków, Poland

[3] Jagiellonian University, Faculty of Physics, Astronomy and Applied Computer Science, M. Smoluchowski Institute of Physics, Łojasiewicza 11, PL-30-348 Kraków, Poland

[4] Institute of Chemistry, Military University of Technology, Kaliskiego 2, PL-00908 Warsaw, Poland

[*]corresponding author, aleksandra.deptuch@ifj.edu.pl

# Electronic Supplementary Information

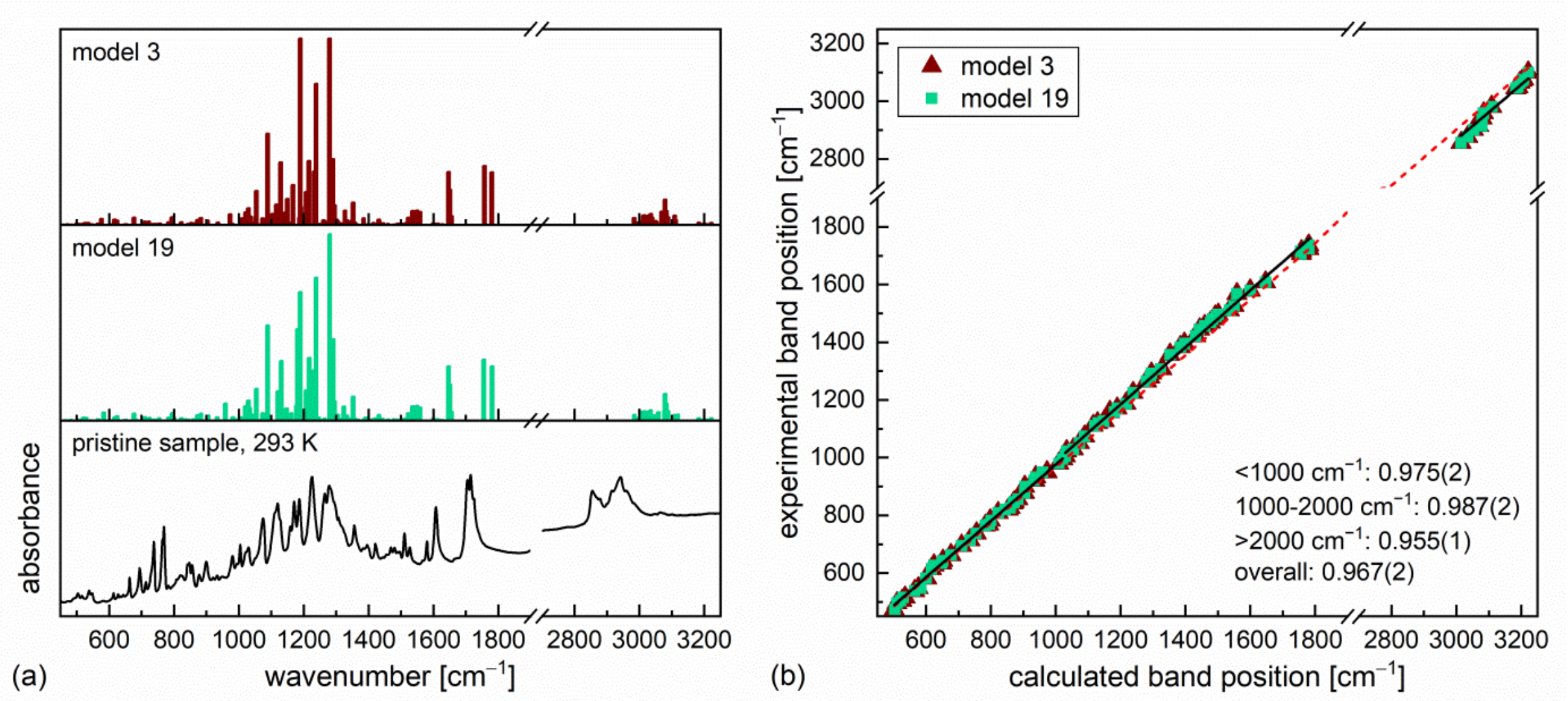

Figure S1. IR spectrum of 7HH6 collected for the pristine sample at 293 K compared to the unscaled calculated spectra for molecular models 3 and 19 (DFT, def2TZVPP basis set, B3LYP-D3(BJ) functional) (a) and scaling between the calculated and experimental band positions (b). The overall scaling factor is equal to 0.967(2). The scaling factors calculated separately for >1000 cm$^{-1}$, 1000-2000 cm$^{-1}$ and >2000 cm$^{-1}$, are equal to 0.975(2), 0.987(2) and 0.955(1), respectively.

Table S1. The band assignment of the experimental IR spectra of 7HH6 measured for the pristine sample at 293 K, based on the DFT calculations (def2TZVPP basis set, B3LYP-D3(BJ) functional). Experimental and calculated band positions are given in $cm^{-1}$. The intra-molecular vibrations are denoted as follows: β – in-plane deformation, γ – out-of-plane deformation, δ – scissoring, ν – stretching, ρ – rocking, τ – twisting, ω – wagging.

| experimental | model 3 | model 19 | interpretation |
|---|---|---|---|
| 475 | 502.1 | 503.7 | $\rho CF_2$(a), δCCC(c) |
| 496 | 510.4 | 511.0 | γPh(d,f,g), δCCC(j) |
| 504 | 526.1 | 521.8 | $\delta CF_2$(a), $\rho CH_2$(b)$^{model3}$, γPh(d)$^{model19}$ |
| 516 | 535.8 | 529.5 | $\delta CF_2$(a), δCOC(b,c), δCCC(c), γPh(f,g), δOCC(b,c)$^{model3}$, $\beta_{asym}$Ph(d)$^{model3}$, γPh(d)$^{model19}$ |
| 538 | 565.4 | 565.1 | γPh(f,g) |
| 548 | 576.2 | 584.0 | $\delta CF_2$(a), δCCO(a,b) |
| 582 | 602.4 | 600.5 | $\rho CF_2$(a), $\rho CH_2$(b) |
| 614 | 616.6 | 617.3 | $\delta CF_2$(a), $\rho CH_2$(b) |
| 627 | 624.7 | 624.7 | $\beta_{asym}$Ph(d,f,g) |
| 634 | 648.7 | 648.8 | $\beta_{asym}$Ph(d,g), γPh(f) |
| 643 | 651.9 | 651.9 | $\beta_{asym}$Ph(d,f,g) |
| 663 | 677.3 | 677.1 | $\beta_{asym}$Ph(d,g), γPh(f) |
| 695 | 709.9 | 710.0 | γPh(d) |
| 714 | 735.6 | 735.9 | $\rho CH_2$(j) |
| 738 | 755.7 | 754.4 | $\rho CH_2$(c) |
| 765 | 783.5 | 783.4 | γPh(d) |
| 769 | 793.8 | 795.2 | γPh(d,f,g), $\rho CH_2$(j) |
| 783 | 798.4 | 800.7 | $\rho CH_2$(j) |
| 809 | 822.7 | 822.9 | γPh(f,g) |
| 820 | 853.6 | 844.2 | $\tau CH_2$(j)$^{model3}$, γPh(f)$^{model19}$ |
| 826 | 859.6 | 860.0 | γPh(f,g) |
| 843 | 869.5 | 869.6 | γPh(d,f,g) |
| 847 | 872.6 | 872.7 | γPh(d) |
| 856 | 883.1 | 883.6 | γPh(f,g) |
| 878 | 904.1 | 904.5 | νCC(a,b)$^{model3}$, $\rho CH_2$(b)$^{model3}$, γPh(f,g)$^{model19}$ |
| 900 | 904.2 | 906.0 | γPh(f,g)$^{model3}$, νCC(a,b)$^{model19}$, $\rho CH_2$(b)$^{model19}$ |
| 921 | 935.8 | 932.9 | $\nu_{asym}$OC*C(h,i,j) |
| 933 | 938.8 | 938.7 | $\tau CH_2$(c), $\rho CH_2$(c) |
| 950 | 973.7 | 959.1 | $\nu_{sym}$COC(b,c), $\nu_{sym}$CCC(c) |
| 980 | 1010.1 | 1009.8 | $\rho CH_2$(b), $\nu_{sym}$CCC(c) |
| 995 | 1021.7 | 1020.8 | $\rho CH_2$(b), $\nu_{sym}$CCC(c) |
| 1005 | 1029.8 | 1029.7 | $\beta_{asym}$Ph(d) |
| 1023 | 1033.8 | 1033.6 | $\beta_{asym}$Ph(d,f,g) |
| 1029 | 1054.4 | 1054.6 | $\nu_{asym}$CCC(c), νCO(c), $\beta_{asym}$Ph(d,f,g) |
| 1051 | 1073.5 | 1073.5 | $\nu_{asym}$CCC(c), $\nu_{sym}$CCC(c) |
| 1074 | 1089.4 | 1089.3 | $\beta_{asym}$Ph(d,f), νCO(e) |
| 1111 | 1115.6 | 1121.2 | $\nu_{sym}CF_2$(a), $\tau CH_2$(b) |
| 1120 | 1129.3 | 1131.6 | $\beta_{asym}$Ph(g), $\nu_{asym}$COC(h,i) |
| 1127 | 1149.8 | 1147.9 | $\nu_{asym}CF_2$(a), $\nu_{sym}$CCC(c), $\beta_{sym}$Ph(d), $\nu_{asym}$COC(b,c)$^{model3}$, $\rho CH_2$(b)$^{model19}$ |
| 1159 | 1167.1 | 1181.5 | $\nu_{asym}CF_2$(a), $\rho CH_2$(b), $\nu_{asym}$COC(b,c)$^{model19}$ |
| 1171 | 1189.5 | 1189.5 | $\beta_{sym}$Ph(d,f) |
| 1187 | 1216.6 | 1217.8 | $\omega CF_2$(a), $\omega CH_2$(b) |
| 1225 | 1238.5 | 1238.5 | $\beta_{sym}$Ph(d,g), $\beta_{asym}$Ph(f), $\nu_{asym}$COC(e,f) |
| 1265 | 1280.4 | 1280.5 | $\beta_{asym}$Ph(d,f), $\nu_{asym}$CCO(d,e) |
| 1279 | 1290.7 | 1291.7 | $\beta_{asym}$Ph(g), $\nu_{asym}$CCO(g,h), δOC*H(h,i) |
| 1292 | 1295.5 | 1295.5 | $\omega CH_2$(c), $\beta_{asym}$Ph(d), $\nu_{asym}$CCO(d,e) |

Table S1 (continuation).

| | | | |
|---|---|---|---|
| 1307 | 1327.7 | 1323.9 | $\omega CF_2$(a), $\tau CH_2$(b,c) |
| 1357 | 1352.6 | 1353.0 | $\tau CH_2$(b), $\omega CH_2$(c), $\beta_{asym}$Ph(d), $\tau CH_2(c)^{model19}$ |
| 1384 | 1385.1 | 1385.1 | $\omega CH_2$(i,j) |
| 1396 | 1397.3 | 1397.5 | $\omega CH_2$(b,c) |
| 1422 | 1432.0 | 1432.3 | $\omega CH_2$(c), $\beta_{asym}$Ph(d) |
| 1445 | 1444.5 | 1443.0 | $\omega CH_2$(b,c) |
| 1457 | 1459.8 | 1459.7 | $\beta_{asym}$Ph(d,f,g) |
| 1469 | 1477.9 | 1478.2 | $\delta CH_2(i,j)^{model3}$, $\delta CH_2(b,c)^{model19}$, $\omega CH_2(c)^{model19}$ |
| 1480 | 1488.2 | 1484.3 | $\delta CH_2(b,c)^{model3}$, $\omega CH_2(c)^{model3}$, $\delta CH_2(i,j)^{model19}$ |
| 1490 | 1491.8 | 1494.5 | $\delta CH_2$(i,j) |
| 1496 | 1501.1 | 1501.1 | $\delta CH_3$(j) |
| 1511 | 1536.4 | 1536.5 | $\beta_{asym}$Ph(f,g) |
| 1527 | 1550.0 | 1550.1 | $\beta_{asym}$Ph(d) |
| 1568 | 1559.3 | 1559.3 | $\beta_{asym}$Ph(f,g) |
| 1580 | 1600.3 | 1600.3 | $\beta_{asym}$Ph(f,g) |
| 1608 | 1647.5 | 1647.5 | $\beta_{sym}$Ph(d,f,g) |
| 1706 | 1758.2 | 1756.1 | $\nu$C=O(h) |
| 1716 | 1758.2 | 1756.1 | $\nu$C=O(h) |
| 1724 | 1780.8 | 1781.2 | $\nu$C=O(e) |
| 1736 | 1780.8 | 1781.2 | $\nu$C=O(e) |
| 2856 | 3014.7 | 3014.4 | $\nu_{sym}CH_2$(c), $\nu_{asym}CH_2$(j) |
| 2877 | 3035.5 | 3034.9 | $\nu_{sym}CH_2$(c), $\nu_{sym}CH_2(b)^{model3}$, $\nu_{asym}CH_2(c)^{model19}$ |
| 2901 | 3053.3 | 3053.4 | $\nu_{asym}CH_2$(c) |
| 2915 | 3069.5 | 3078.1 | $\nu_{asym}$CH(i,j) |
| 2941 | 3079.5 | 3079.4 | $\nu_{asym}CH_2$(c) |
| 2959 | 3083.7 | 3083.3 | $\nu_{asym}CH_2$(j) |
| 2981 | 3109.1 | 3110.0 | $\nu CH_3$(i) |
| 3046 | 3183.6 | 3183.6 | $\nu_{asym}$CH(f,g) |
| 3049 | 3189.9 | 3190.0 | $\nu_{asym}$CH(f,g) |
| 3065 | 3196.4 | 3196.4 | $\nu_{asym}$CH(d) |
| 3072 | 3202.8 | 3202.9 | $\nu_{sym}$CH(f), $\nu_{asym}$CH(d,g) |
| 3076 | 3208.1 | 3207.6 | $\nu_{sym}$CH(f,g) |
| 3100 | 3221.8 | 3222.3 | $\nu_{sym}$CH(d) |

Table S2. Atomic coordinates of the molecular models 3 and 19 of 7HH6 optimized by the DFT method (def2TZVPP basis set, B3LYP-D3(BJ) functional).

| atom | model 3 | | | model 19 | | |
|---|---|---|---|---|---|---|
| | *x* | *y* | *z* | *x* | *y* | *z* |
| C | 13.48027 | -0.39029 | -1.14851 | 14.15474 | -2.46000 | -1.39879 |
| C | 12.25471 | -0.05772 | -1.70331 | 13.22369 | -2.54803 | -2.42131 |
| C | 11.20606 | -0.96247 | -1.63815 | 12.21560 | -3.49731 | -2.34865 |
| C | 11.36263 | -2.20340 | -1.01452 | 12.11842 | -4.36320 | -1.25589 |
| C | 12.61158 | -2.50999 | -0.46402 | 13.07310 | -4.24963 | -0.23967 |
| C | 13.66936 | -1.61733 | -0.52937 | 14.08939 | -3.31009 | -0.30441 |
| H | 12.13157 | 0.90728 | -2.17405 | 13.29275 | -1.86944 | -3.25960 |
| H | 10.24592 | -0.68907 | -2.05265 | 11.47881 | -3.54692 | -3.13804 |
| H | 12.76636 | -3.47332 | 0.00140 | 13.03557 | -4.92425 | 0.60415 |
| H | 14.62869 | -1.86801 | -0.10422 | 14.82190 | -3.23557 | 0.48424 |
| C | 9.34205 | -3.30400 | -2.00078 | 10.59195 | -6.03733 | -2.32101 |
| C | 10.24279 | -3.16248 | -0.93840 | 11.03995 | -5.36828 | -1.17571 |
| C | 10.05091 | -3.95422 | 0.19928 | 10.43563 | -5.67750 | 0.04794 |
| C | 8.29113 | -4.20066 | -1.92849 | 9.57975 | -6.97734 | -2.24631 |
| H | 9.48433 | -2.72158 | -2.90041 | 11.06073 | -5.83396 | -3.27374 |
| C | 8.10932 | -4.98511 | -0.78964 | 8.98522 | -7.27864 | -1.02115 |
| H | 7.60108 | -4.31500 | -2.75188 | 9.23984 | -7.49872 | -3.12946 |
| C | 9.00055 | -4.85342 | 0.27553 | 9.42329 | -6.61895 | 0.12741 |
| H | 10.71800 | -3.84343 | 1.04277 | 10.74622 | -5.15435 | 0.94158 |
| H | 8.86003 | -5.45021 | 1.16368 | 8.96083 | -6.84034 | 1.07718 |
| C | 6.96139 | -5.92899 | -0.76050 | 7.89925 | -8.29275 | -0.99448 |
| O | 6.17266 | -6.05769 | -1.66769 | 7.48792 | -8.86947 | -1.97490 |
| O | 6.90449 | -6.61809 | 0.39450 | 7.42544 | -8.49724 | 0.24761 |
| C | 5.80740 | -7.55063 | 0.59613 | 6.32684 | -9.43800 | 0.40442 |
| C | 4.88256 | -6.97513 | 1.65845 | 5.00466 | -8.74656 | 0.08204 |
| C | 6.42209 | -8.87983 | 0.98908 | 6.43594 | -9.97391 | 1.81834 |
| H | 5.27812 | -7.63877 | -0.35081 | 6.48388 | -10.23962 | -0.31591 |
| H | 6.99563 | -8.78269 | 1.91180 | 6.37861 | -9.17107 | 2.55205 |
| H | 7.08600 | -9.24446 | 0.20603 | 7.38194 | -10.49597 | 1.95661 |
| H | 5.63884 | -9.62161 | 1.14797 | 5.62402 | -10.67546 | 2.01229 |
| C | 4.23880 | -5.64694 | 1.27202 | 4.67737 | -7.53045 | 0.94548 |
| H | 5.44540 | -6.85978 | 2.58980 | 4.21099 | -9.49358 | 0.17668 |
| H | 4.10304 | -7.71705 | 1.85487 | 5.02684 | -8.45471 | -0.96930 |
| C | 3.29017 | -5.11254 | 2.34116 | 3.40168 | -6.82164 | 0.49804 |
| H | 3.69973 | -5.76530 | 0.32807 | 5.51274 | -6.82680 | 0.91004 |
| H | 5.01972 | -4.90601 | 1.08453 | 4.57038 | -7.82823 | 1.99187 |
| C | 2.65308 | -3.77700 | 1.96843 | 3.05895 | -5.60084 | 1.34675 |
| H | 3.83351 | -5.00255 | 3.28558 | 2.56474 | -7.52744 | 0.52454 |
| H | 2.50196 | -5.84910 | 2.52968 | 3.50694 | -6.51714 | -0.54815 |
| C | 1.70335 | -3.23573 | 3.03401 | 1.78702 | -4.88478 | 0.89947 |
| H | 2.11108 | -3.88621 | 1.02333 | 3.89711 | -4.89657 | 1.32122 |
| H | 3.44158 | -3.04099 | 1.78010 | 2.95215 | -5.90515 | 2.39341 |
| C | 1.07305 | -1.89994 | 2.65039 | 1.45497 | -3.66505 | 1.75425 |
| H | 0.49780 | -1.98881 | 1.72652 | 2.26234 | -2.93065 | 1.71990 |
| H | 1.83796 | -1.13762 | 2.48873 | 1.31079 | -3.94598 | 2.79961 |
| H | 0.39977 | -1.53572 | 3.42750 | 0.54318 | -3.17252 | 1.41384 |
| H | 2.24660 | -3.12707 | 3.97755 | 0.95025 | -5.58926 | 0.92512 |
| H | 0.91602 | -3.97215 | 3.22112 | 1.89510 | -4.58126 | -0.14604 |
| C | 16.29547 | 1.88753 | -0.60876 | 16.59578 | 0.22584 | -0.91917 |
| C | 17.09176 | 2.42901 | 0.39814 | 16.96236 | 1.22797 | -0.02336 |
| C | 18.11372 | 3.31697 | 0.10311 | 18.03813 | 2.06178 | -0.28286 |
| H | 16.89740 | 2.14313 | 1.42188 | 16.38910 | 1.34542 | 0.88501 |
| C | 18.35121 | 3.67556 | -1.22660 | 18.76909 | 1.89511 | -1.46229 |
| H | 18.71392 | 3.72066 | 0.90294 | 18.29900 | 2.82914 | 0.42850 |
| C | 17.55650 | 3.13521 | -2.24393 | 18.40735 | 0.89071 | -2.36721 |
| C | 16.54180 | 2.25313 | -1.93782 | 17.33422 | 0.06761 | -2.09832 |
| H | 17.75775 | 3.42560 | -3.26521 | 18.98639 | 0.77907 | -3.27278 |
| H | 15.93191 | 1.83993 | -2.72646 | 17.06000 | -0.70541 | -2.79967 |
| O | 19.31879 | 4.53114 | -1.62443 | 19.83363 | 2.65038 | -1.81269 |
| C | 20.16785 | 5.12472 | -0.64272 | 20.26422 | 3.69245 | -0.93757 |
| C | 21.14566 | 6.03368 | -1.35661 | 21.46444 | 4.36711 | -1.56697 |
| H | 20.69721 | 4.34026 | -0.09281 | 20.52434 | 3.26848 | 0.03739 |
| H | 19.56226 | 5.69102 | 0.07180 | 19.44921 | 4.40806 | -0.79033 |
| C | 22.10383 | 6.72144 | -0.38753 | 22.00482 | 5.50282 | -0.70187 |
| H | 20.58404 | 6.78090 | -1.92198 | 21.17976 | 4.74776 | -2.55058 |
| H | 21.70785 | 5.44468 | -2.08485 | 22.24309 | 3.61890 | -1.73204 |
| C | 23.09579 | 7.64554 | -1.08841 | 23.21793 | 6.19373 | -1.31849 |
| H | 22.65524 | 5.96563 | 0.18080 | 22.27495 | 5.11455 | 0.28532 |
| H | 21.53141 | 7.29756 | 0.34654 | 21.21561 | 6.24247 | -0.53315 |
| C | 24.05851 | 8.33631 | -0.12659 | 23.76672 | 7.32866 | -0.45832 |
| H | 22.54607 | 8.40237 | -1.65628 | 22.94958 | 6.58379 | -2.30506 |
| H | 23.66864 | 7.07128 | -1.82278 | 24.00715 | 5.45506 | -1.48822 |
| O | 26.87310 | 10.77554 | -0.62040 | 26.66455 | 9.68483 | -0.86281 |
| C | 27.87583 | 11.41376 | 0.11610 | 27.17984 | 10.85525 | -0.30537 |
| C | 29.27103 | 10.86672 | -0.18934 | 28.04598 | 10.64876 | 0.93920 |
| H | 27.70967 | 11.35955 | 1.19381 | 26.39392 | 11.56002 | -0.00798 |
| H | 27.92075 | 12.46685 | -0.17232 | 27.80603 | 11.32787 | -1.05989 |
| C | 29.52011 | 9.39026 | 0.22097 | 29.15947 | 9.57586 | 0.77864 |
| F | 30.16980 | 11.64153 | 0.48662 | 27.28488 | 10.29951 | 2.01277 |
| F | 29.53850 | 10.98264 | -1.51201 | 28.63522 | 11.84205 | 1.23552 |
| C | 31.00061 | 8.90442 | 0.18086 | 30.28687 | 9.60526 | 1.85476 |
| F | 28.80854 | 8.57763 | -0.58756 | 29.75383 | 9.74227 | -0.42489 |
| F | 29.08177 | 9.21635 | 1.49410 | 28.59797 | 8.34985 | 0.82055 |
| F | 31.02771 | 7.58801 | 0.41838 | 31.06462 | 8.52905 | 1.69910 |
| F | 31.73300 | 9.51366 | 1.11320 | 29.76780 | 9.58020 | 3.08490 |
| F | 31.54355 | 9.13127 | -1.01594 | 31.04615 | 10.69368 | 1.72689 |
| C | 15.22561 | 0.94879 | -0.21978 | 15.44090 | -0.62563 | -0.57525 |
| O | 14.51597 | 0.51890 | -1.31160 | 15.18798 | -1.54736 | -1.55874 |
| O | 14.98281 | 0.59582 | 0.90450 | 14.78526 | -0.53845 | 0.42974 |
| C | 25.04679 | 9.25849 | -0.83580 | 24.98300 | 8.01038 | -1.07974 |
| H | 24.60804 | 7.57753 | 0.43952 | 24.03227 | 6.93671 | 0.52850 |
| H | 23.48417 | 8.90921 | 0.60853 | 22.97697 | 8.06799 | -0.29031 |
| C | 26.00055 | 9.92913 | 0.12894 | 25.52201 | 9.12657 | -0.21050 |
| H | 24.50862 | 10.02890 | -1.39303 | 24.72647 | 8.41650 | -2.06104 |
| H | 25.63090 | 8.69248 | -1.56491 | 25.77896 | 7.27935 | -1.23974 |
| H | 26.57596 | 9.17778 | 0.67738 | 25.79592 | 8.74336 | 0.77563 |
| H | 25.45100 | 10.53233 | 0.86200 | 24.76828 | 9.91103 | -0.06853 |

Table S3. Theoretical IR spectra for models 3 and 19 calculated by the DFT method (def2TZVPP basis set, B3LYP-D3(BJ) functional).

| model 3 | | model 19 | |
|---|---|---|---|
| wavenumber [$cm^{-1}$] | intensity | wavenumber [$cm^{-1}$] | intensity |
| 3.385 | 0.0056939 | 3.406 | 0.00347898 |
| 4.0246 | 0.0102448 | 3.838 | 0.0113887 |
| 6.0017 | 0.011954 | 6.7639 | 0.0176028 |
| 9.1439 | 0.00642033 | 9.4861 | 0.00805198 |
| 10.6053 | 0.0155861 | 11.4963 | 0.00507625 |
| 11.6975 | 0.00521318 | 13.9184 | 0.0128875 |
| 17.067 | 0.0116549 | 17.9736 | 0.0109621 |
| 17.8812 | 0.0242071 | 18.8625 | 0.00749403 |
| 24.0242 | 0.00217928 | 26.5919 | 0.0044089 |
| 28.2331 | 0.0463311 | 28.6858 | 0.0245388 |
| 35.679 | 0.00799069 | 33.8561 | 0.00775659 |
| 41.4378 | 0.0466729 | 39.6805 | 0.0235761 |
| 43.2704 | 0.0393339 | 41.7493 | 0.0783427 |
| 46.0571 | 0.0975228 | 49.256 | 0.0478415 |
| 49.7723 | 0.0808256 | 53.0226 | 0.166214 |
| 53.4399 | 0.118685 | 57.1417 | 0.00495591 |
| 58.5035 | 0.0170497 | 62.3157 | 0.0631905 |
| 59.3322 | 0.00930467 | 62.6007 | 0.00904754 |
| 68.56 | 0.0127018 | 67.3864 | 0.00437608 |
| 71.8853 | 0.0346014 | 73.9649 | 0.015371 |
| 75.8697 | 0.0379985 | 77.1271 | 0.0211693 |
| 77.202 | 0.00664467 | 78.7037 | 0.0994463 |
| 85.9191 | 0.021013 | 83.2285 | 0.0213553 |
| 89.9985 | 0.0388211 | 83.9736 | 0.00588582 |
| 94.9966 | 0.0711577 | 100.837 | 0.0395707 |
| 109.002 | 0.0294203 | 105.955 | 0.0552917 |
| 110.259 | 0.0689144 | 116.275 | 0.0291337 |
| 115.14 | 0.0393125 | 120.263 | 0.0328425 |
| 124.427 | 0.0664894 | 123.546 | 0.010054 |
| 129.238 | 0.0218462 | 127.008 | 0.0474148 |
| 129.89 | 0.129133 | 138.049 | 0.171498 |
| 144.165 | 0.245799 | 145.525 | 0.043345 |
| 151.191 | 0.0426776 | 151.536 | 0.0418572 |
| 158.74 | 0.0226047 | 157.233 | 0.00506531 |
| 160.868 | 0.192311 | 165.878 | 0.0787584 |
| 168.144 | 0.228183 | 166.745 | 0.583156 |
| 169.787 | 0.0487347 | 171.344 | 0.21445 |
| 174.213 | 0.0344839 | 173.939 | 0.0274161 |
| 181.271 | 0.0520463 | 198.738 | 0.0326783 |
| 191.865 | 0.139218 | 200.416 | 0.294565 |
| 214.226 | 0.136568 | 206.822 | 0.121349 |
| 220.936 | 0.21858 | 221.157 | 0.139422 |
| 229.674 | 0.0711471 | 234.672 | 0.0732118 |
| 238.298 | 0.012189 | 240.06 | 0.015557 |
| 241.606 | 0.0354988 | 243.869 | 0.0642846 |
| 245.283 | 0.0541401 | 246.089 | 0.00510907 |
| 246.224 | 0.0117297 | 251.559 | 0.542338 |
| 257.168 | 0.61679 | 262.135 | 0.0681027 |
| 278.753 | 0.533443 | 279.211 | 0.63254 |
| 286.244 | 0.690105 | 286.538 | 0.885182 |
| 292.38 | 0.24096 | 292.88 | 0.0471194 |
| 297.044 | 0.655151 | 298.269 | 0.51595 |
| 305.262 | 0.225833 | 304.963 | 0.177603 |
| 314.973 | 0.147871 | 323.041 | 0.0760015 |
| 328.971 | 0.0919891 | 336.072 | 0.15872 |
| 341.96 | 0.12486 | 347.394 | 0.288099 |
| 350.81 | 0.463343 | 351.124 | 0.781228 |
| 363.033 | 0.199393 | 367.061 | 0.0657396 |
| 374.216 | 0.014379 | 373.9 | 0.0289368 |
| 386.922 | 0.193294 | 393.398 | 0.0727523 |
| 393.766 | 0.513423 | 394.494 | 0.752652 |
| 406.779 | 0.339733 | 405.01 | 0.170295 |
| 411.845 | 0.0912627 | 409.692 | 0.525457 |
| 421.465 | 0.0144217 | 418.329 | 0.379121 |
| 422.062 | 0.74353 | 421.47 | 0.0271536 |
| 428.297 | 0.0184811 | 428.308 | 0.0231057 |
| 429.612 | 0.00300185 | 429.724 | 0.00402599 |
| 436.372 | 0.516201 | 436.036 | 0.748375 |
| 457.425 | 0.0488735 | 443.559 | 0.407982 |
| 471.016 | 0.297354 | 461.286 | 0.083222 |
| 483.217 | 0.0933351 | 471.417 | 0.127639 |
| 499.82 | 0.690938 | 483.401 | 0.277662 |
| 502.121 | 0.455373 | 503.722 | 0.23354 |
| 510.442 | 0.547875 | 510.964 | 0.417773 |
| 519.46 | 0.405891 | 521.333 | 1.61402 |
| 522.29 | 0.686697 | 521.808 | 1.1542 |
| 526.145 | 1.09677 | 529.525 | 1.32779 |
| 533.678 | 0.367796 | 530.274 | 0.353631 |
| 535.796 | 0.962312 | 542.527 | 0.208498 |
| 546.531 | 0.258437 | 565.126 | 0.617049 |
| 565.36 | 0.693791 | 583.963 | 4.3128 |
| 576.196 | 3.19057 | 600.511 | 0.293416 |
| 602.362 | 0.360126 | 617.34 | 1.71058 |
| 616.62 | 2.88787 | 624.747 | 2.29508 |
| 624.712 | 2.35204 | 628.183 | 0.235028 |
| 648.654 | 0.335833 | 648.757 | 0.356486 |
| 651.92 | 0.271598 | 651.936 | 0.28474 |
| 659.982 | 0.0403487 | 660.01 | 0.0381813 |
| 675.741 | 0.10217 | 677.076 | 3.66055 |
| 677.306 | 3.96586 | 683.584 | 0.502034 |
| 709.872 | 1.69099 | 709.984 | 1.73634 |
| 721.729 | 1.81838 | 724.359 | 1.41636 |
| 735.589 | 0.478747 | 735.852 | 0.546769 |
| 739.209 | 0.41526 | 739.318 | 0.426755 |
| 742.643 | 0.502709 | 742.916 | 0.476511 |
| 748.176 | 0.0432437 | 749.403 | 0.15418 |
| 754.616 | 0.598939 | 754.356 | 1.23748 |
| 755.655 | 0.688663 | 755.561 | 0.234175 |
| 755.927 | 0.62963 | 756.056 | 0.520895 |
| 783.472 | 1.47108 | 783.396 | 1.57648 |
| 791.441 | 0.104467 | 791.964 | 0.0309826 |
| 793.789 | 4.15137 | 795.179 | 3.94439 |
| 798.408 | 0.73242 | 800.686 | 1.46755 |
| 822.723 | 1.30485 | 822.909 | 1.19783 |
| 833.761 | 0.0249228 | 834.046 | 0.0361464 |
| 839.42 | 0.193881 | 834.875 | 0.60752 |
| 843.056 | 0.559092 | 839.806 | 0.120736 |
| 853.547 | 1.01733 | 844.22 | 0.575126 |
| 854.232 | 0.149548 | 854.551 | 0.179627 |
| 859.579 | 0.869533 | 859.998 | 0.652637 |
| 864.335 | 0.278478 | 864.714 | 0.744567 |
| 869.489 | 0.439061 | 869.561 | 0.84571 |
| 872.559 | 2.68508 | 872.717 | 2.86076 |
| 883.11 | 3.65356 | 883.566 | 3.63828 |
| 895.913 | 0.150637 | 890.586 | 0.00850053 |
| 901.19 | 0.110983 | 900.466 | 0.105201 |
| 904.09 | 2.19258 | 904.476 | 1.03419 |
| 904.204 | 0.69159 | 905.957 | 2.56167 |
| 935.762 | 1.26764 | 932.912 | 1.66935 |
| 938.779 | 0.0982385 | 938.672 | 0.127213 |
| 973.652 | 5.61879 | 959.14 | 9.13671 |
| 976.921 | 0.123225 | 977.197 | 0.190337 |
| 980.813 | 0.527909 | 981.397 | 1.76147 |
| 986.025 | 0.00424105 | 986.045 | 0.0203706 |
| 993.542 | 0.0417909 | 992.009 | 0.630275 |
| 999.17 | 0.0288861 | 993.747 | 0.0317484 |
| 1003.25 | 0.0116015 | 999.223 | 0.028346 |
| 1010.13 | 3.48683 | 1002.93 | 0.0216178 |
| 1012.5 | 0.0770333 | 1009.77 | 1.79667 |
| 1012.99 | 1.58117 | 1013.16 | 0.117115 |
| 1021.72 | 6.69011 | 1017.66 | 0.0483338 |
| 1025.3 | 4.69338 | 1020.82 | 7.61671 |
| 1028.9 | 0.0411927 | 1028.74 | 0.0251406 |
| 1029.77 | 8.85293 | 1029.7 | 10.5932 |
| 1033.81 | 4.2705 | 1033.62 | 6.81405 |
| 1035.71 | 4.35544 | 1034.07 | 3.57202 |
| 1042.72 | 0.485776 | 1039.89 | 3.13547 |
| 1044.91 | 1.02238 | 1042.73 | 0.175929 |
| 1052.66 | 4.36977 | 1052.63 | 2.79733 |
| 1054.42 | 18.2115 | 1054.61 | 16.9121 |
| 1065.71 | 0.0634662 | 1060.88 | 0.0517033 |
| 1069.7 | 0.522525 | 1069.39 | 0.402708 |
| 1073.46 | 3.60196 | 1072.64 | 0.00748309 |
| 1074.4 | 0.35471 | 1073.5 | 3.84613 |
| 1084.94 | 2.40128 | 1084.75 | 1.5376 |
| 1089.36 | 48.8441 | 1089.28 | 51.0109 |
| 1103.47 | 5.80831 | 1101.25 | 0.0442859 |
| 1115.58 | 10.8205 | 1121.21 | 15.6356 |
| 1128.08 | 5.94159 | 1127.97 | 2.28746 |
| 1129.27 | 33.5435 | 1131.61 | 31.9143 |
| 1136.65 | 0.631404 | 1136.55 | 0.584917 |
| 1142.71 | 6.63373 | 1143.36 | 6.24668 |
| 1142.86 | 5.89336 | 1143.71 | 3.77056 |
| 1143.81 | 0.356398 | 1147.86 | 6.43443 |
| 1149.77 | 13.8567 | 1149.56 | 0.525829 |
| 1150.6 | 1.75313 | 1162.03 | 3.18992 |
| 1157.37 | 0.868678 | 1162.08 | 3.92977 |
| 1167.05 | 21.4272 | 1178.83 | 7.95065 |
| 1187.42 | 3.97639 | 1181.47 | 48.9978 |
| 1189.54 | 99.9861 | 1188.97 | 26.1868 |
| 1191.77 | 9.68667 | 1189.54 | 68.9113 |
| 1196.45 | 0.483223 | 1191.84 | 2.00254 |
| 1196.53 | 0.323997 | 1196.51 | 0.282848 |
| 1207.73 | 17.6005 | 1208.17 | 16.1404 |
| 1208.72 | 0.845689 | 1208.39 | 1.0523 |
| 1216.59 | 34.2453 | 1217.78 | 33.5312 |
| 1232.82 | 0.799539 | 1225.73 | 26.8041 |
| 1233.07 | 28.7272 | 1231.56 | 1.04178 |
| 1238.48 | 75.6705 | 1238.46 | 76.4962 |
| 1247.37 | 0.533849 | 1246.73 | 1.51504 |
| 1256.4 | 0.156748 | 1255.97 | 0.361792 |
| 1261.7 | 2.50932 | 1262.33 | 0.306424 |
| 1264.71 | 1.92944 | 1264.65 | 1.08073 |
| 1280.36 | 100 | 1280.47 | 100 |
| 1287.31 | 1.05552 | 1287.19 | 0.0738682 |
| 1290.71 | 35.3075 | 1291.73 | 43.4826 |
| 1295.5 | 10.5544 | 1295.54 | 10.3468 |
| 1298.66 | 0.415302 | 1299.02 | 1.06931 |
| 1301.88 | 2.89944 | 1301.77 | 3.13711 |
| 1305.46 | 1.93871 | 1305.62 | 3.02241 |
| 1315.02 | 0.801099 | 1314.9 | 0.83571 |
| 1315.39 | 0.378671 | 1321.96 | 0.554744 |
| 1325.13 | 0.127146 | 1323.88 | 7.37664 |
| 1326.05 | 0.035424 | 1325.5 | 1.7446 |
| 1326.77 | 0.157271 | 1325.73 | 2.59476 |
| 1327.65 | 7.67199 | 1326.25 | 0.025239 |
| 1332.97 | 2.42559 | 1332.93 | 2.63294 |
| 1336.15 | 0.109349 | 1335.74 | 0.134553 |
| 1337.04 | 0.514887 | 1337.07 | 0.684364 |
| 1339.62 | 0.0229679 | 1339.62 | 0.0427871 |
| 1342.4 | 0.202502 | 1342.12 | 0.978075 |
| 1342.9 | 0.123193 | 1342.4 | 0.37492 |
| 1344.85 | 0.338771 | 1342.67 | 0.0594599 |
| 1345.22 | 2.06495 | 1347.23 | 1.10208 |
| 1347.26 | 0.628637 | 1349.32 | 1.07588 |
| 1352.57 | 11.9093 | 1353.02 | 12.8522 |
| 1358.33 | 1.24637 | 1358.18 | 0.992144 |
| 1362.75 | 0.355062 | 1367.65 | 0.551057 |
| 1385.14 | 3.59242 | 1385.06 | 0.710773 |
| 1397.32 | 0.239956 | 1392.15 | 1.59757 |
| 1397.84 | 0.162784 | 1397.45 | 0.217053 |
| 1406.65 | 0.126922 | 1404.21 | 0.12241 |
| 1410.95 | 0.293872 | 1410.99 | 0.104315 |
| 1412.43 | 0.481749 | 1412.69 | 0.610616 |
| 1415.85 | 1.07073 | 1415.52 | 0.875915 |
| 1417.21 | 0.317438 | 1417.35 | 0.161871 |
| 1418.1 | 0.126409 | 1423.43 | 1.2308 |
| 1431.95 | 2.85164 | 1432.25 | 3.0177 |
| 1432.8 | 1.85124 | 1432.88 | 1.96585 |
| 1444.51 | 0.810489 | 1443.01 | 0.850786 |
| 1459.77 | 0.490509 | 1459.72 | 0.514659 |
| 1460.35 | 0.298679 | 1460.4 | 0.345929 |
| 1477.86 | 0.050978 | 1478.18 | 0.725586 |
| 1487.99 | 0.033437 | 1484.29 | 0.434741 |
| 1488.15 | 0.53653 | 1487.85 | 0.0446797 |
| 1489.4 | 0.0554541 | 1489.8 | 0.00147693 |
| 1489.84 | 0.00916579 | 1490.33 | 0.00377437 |
| 1491.27 | 0.0425066 | 1491.16 | 0.0122311 |
| 1491.77 | 0.671474 | 1494.5 | 0.518707 |
| 1497.86 | 0.0840198 | 1497.79 | 0.04701 |
| 1497.91 | 0.180314 | 1499.39 | 0.148918 |
| 1500.29 | 0.621041 | 1501.14 | 0.749491 |
| 1501.12 | 0.746531 | 1505.21 | 0.197689 |
| 1505.36 | 0.170176 | 1505.98 | 0.8648 |
| 1508.34 | 0.0539692 | 1510.32 | 0.0356431 |
| 1514.2 | 0.670214 | 1513.99 | 0.575684 |
| 1516.81 | 1.07116 | 1518.18 | 1.20426 |
| 1523.2 | 3.85039 | 1523.21 | 3.9513 |
| 1528.66 | 0.565224 | 1527.37 | 0.681705 |
| 1536.39 | 7.48593 | 1536.54 | 7.59961 |
| 1550.04 | 7.60866 | 1550.09 | 7.79398 |
| 1559.27 | 6.97204 | 1559.3 | 6.94699 |
| 1600.3 | 0.876049 | 1600.33 | 0.945364 |
| 1613.28 | 0.667661 | 1613.34 | 0.709165 |
| 1626.3 | 0.310953 | 1626.36 | 0.318677 |
| 1647.47 | 28.3529 | 1647.5 | 29.1129 |
| 1651.65 | 18.6867 | 1651.65 | 19.1861 |
| 1656.14 | 4.68372 | 1656.12 | 4.61192 |
| 1758.22 | 31.4954 | 1756.09 | 32.66 |
| 1780.79 | 28.0931 | 1781.16 | 28.984 |
| 2983.85 | 3.64077 | 2984.55 | 2.94433 |
| 2997.31 | 0.200301 | 2997.37 | 0.0738244 |
| 2998.44 | 1.41901 | 2998.57 | 1.57174 |
| 3001.95 | 0.550653 | 3002.07 | 0.976117 |
| 3003.1 | 0.217319 | 3003.24 | 0.196541 |
| 3004.3 | 1.52962 | 3004.19 | 0.551025 |
| 3012.62 | 1.06698 | 3004.52 | 4.82489 |
| 3012.97 | 4.87885 | 3013.01 | 3.70578 |
| 3014.51 | 0.29541 | 3014.43 | 5.00252 |
| 3014.69 | 5.1966 | 3014.99 | 0.962124 |
| 3020.81 | 0.209895 | 3020.75 | 0.400641 |
| 3024.14 | 2.90354 | 3020.8 | 0.576526 |
| 3025.8 | 4.13561 | 3024.09 | 3.64249 |
| 3027.97 | 4.91843 | 3025.82 | 4.85821 |
| 3028.87 | 0.0650472 | 3029.05 | 0.0523269 |
| 3031.73 | 2.26244 | 3030.24 | 3.18975 |
| 3034.48 | 2.51305 | 3032.34 | 1.56355 |
| 3034.56 | 1.19851 | 3034.59 | 2.5957 |
| 3035.54 | 5.86719 | 3034.88 | 5.56337 |
| 3037.76 | 2.76151 | 3044.19 | 2.4791 |
| 3039.3 | 3.89438 | 3044.58 | 2.79592 |
| 3040.92 | 1.14362 | 3046.78 | 0.499759 |
| 3049.6 | 2.15721 | 3053.35 | 1.34525 |
| 3053.34 | 1.48056 | 3058.6 | 4.42615 |
| 3069.51 | 7.0687 | 3076.71 | 0.326346 |
| 3076.96 | 0.139282 | 3078.05 | 5.25754 |
| 3079.52 | 13.56 | 3079.1 | 1.68517 |
| 3083.67 | 7.30525 | 3079.37 | 14.3291 |
| 3086.02 | 1.02818 | 3083.34 | 8.94859 |
| 3088.69 | 1.49794 | 3088.9 | 4.89425 |
| 3088.74 | 4.02751 | 3110.01 | 3.03774 |
| 3109.05 | 4.96179 | 3113.53 | 1.04075 |
| 3112.63 | 2.41089 | 3118.4 | 3.12484 |
| 3182.75 | 1.01481 | 3182.77 | 1.15035 |
| 3183.56 | 1.25941 | 3183.64 | 1.16215 |
| 3187.97 | 0.225107 | 3187.94 | 0.265453 |
| 3189.85 | 0.480008 | 3190.04 | 0.513368 |
| 3196.44 | 0.518807 | 3196.35 | 0.523072 |
| 3201.41 | 0.0152763 | 3201.53 | 0.0116185 |
| 3202.75 | 0.530163 | 3202.9 | 0.452607 |
| 3208.11 | 0.383831 | 3207.63 | 0.411001 |
| 3217.31 | 0.224989 | 3216.32 | 0.22365 |
| 3219.19 | 0.2541 | 3219.29 | 0.267663 |
| 3220.42 | 0.539276 | 3219.56 | 0.549537 |
| 3221.82 | 1.04464 | 3222.27 | 1.1249 |

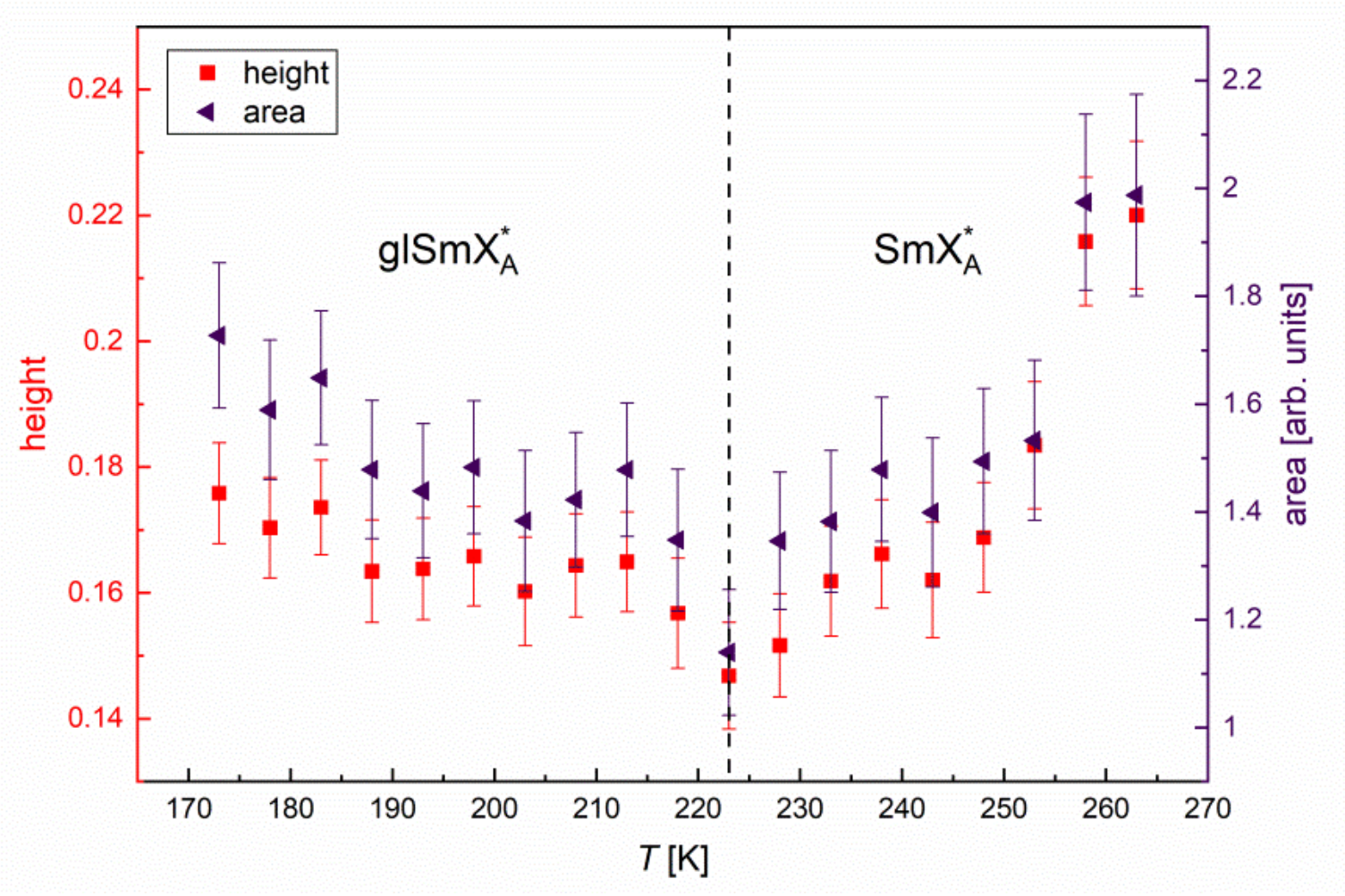

Figure S2. Height and area of the band IV in the SmX$_A$* phase and its glass.

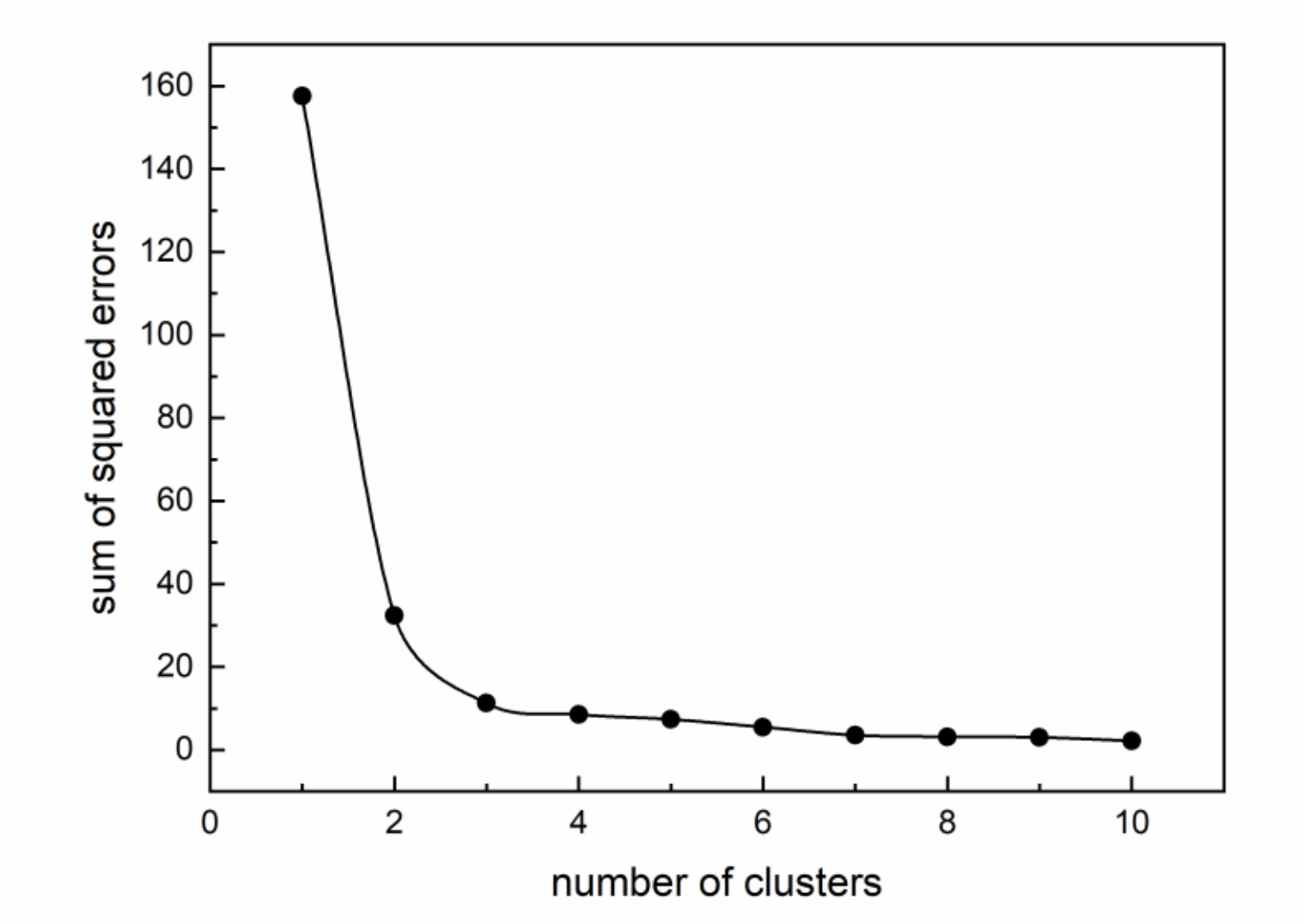

Figure S3. Application of the elbow method for the IR spectra of 7HH6 in the 1650-1850 cm$^{-1}$ range. The line connecting points is a guide to an eye.